\documentclass[journal]{IEEEtran}
\usepackage{cite}
\usepackage{graphicx}
\DeclareGraphicsExtensions{.pdf,.jpeg,.png}
\usepackage{amsmath}
\usepackage{algorithmic}
\usepackage{array}
\usepackage{lineno}
\usepackage{hyperref}
\usepackage{subfigure}
\usepackage{soul}
\usepackage{amssymb}
\usepackage{color}

\modulolinenumbers[5]

\begin{document}
\title{Generalized Debye Sources Based EFIE Solver on Subdivision Surfaces}

\author{Xin~Fu,~\IEEEmembership{Student~Member,~IEEE,}
        Jie~Li,~\IEEEmembership{Student~Member,~IEEE,}
        Li~Jun~Jiang,~\IEEEmembership{Senior~Member,~IEEE}
        Balasubramaniam~Shanker,~\IEEEmembership{Fellow,~IEEE}
\thanks{X. Fu and L. J. Jiang are with the Department of Electrical and Electronic Engineering, The University of Hong Kong, Pokfulam, Hong Kong 999077, China (e-mail: xinfu@eee.hku.hk, jianglj@hku.hk).}
\thanks{J. Li and B. Shanker are with Department of Electrical and Computer Engineering, Michigan State University, East Lansing, MI 48824-1226, USA
(e-mail: jieli@egr.msu.edu, bshanker@egr.msu.edu).}
\thanks{Manuscript received April 19, 2005; revised August 26, 2015.}}

\markboth{Journal of \LaTeX\ Class Files,~Vol.~14, No.~8, August~2015}{Fu \MakeLowercase{\textit{et al.}}: Generalized Debye sources based EFIE solver on subdivision surfaces}

\maketitle

\begin{abstract}
The electric field integral equation is a well known workhorse for obtaining fields scattered by a perfect electric conducting (PEC) object. As a result, the nuances and challenges of solving this equation have been examined for a while. Two recent papers motivate the effort presented in this paper. Unlike traditional work that uses equivalent currents defined on surfaces, recent research proposes a technique that results in well conditioned systems by employing generalized Debye sources (GDS) as unknowns. In a complementary effort, some of us developed a method that exploits the same representation for both the geometry (subdivision surface representations) and functions defined on the geometry, also known as isogeometric analysis (IGA). The challenge in generalizing GDS method to a discretized geometry is the complexity of the intermediate operators. However, thanks to our earlier work on subdivision surfaces, the additional smoothness of geometric representation permits discretizing these intermediate operations. In this paper, we employ both ideas to present a well conditioned GDS-EFIE. Here, the intermediate surface Laplacian is well discretized by using subdivision basis. Likewise, using subdivision basis to represent the sources, results in an efficient and accurate IGA framework. Numerous
results are presented to demonstrate the efficacy of the approach.

\end{abstract}
\begin{IEEEkeywords}
Debye Sources, Electric Field Integral Equation, Surface Laplacian, Subdivision Surfaces, Isogeometric Analysis
\end{IEEEkeywords}

\IEEEpeerreviewmaketitle

\section{Introduction}
Surface integral equation (SIE) solvers have been the mainstay in computational electromagnetics, including a range of problems \cite{poggio1970,peterson1998,jin2011}. In particular, the electric field integral equation (EFIE) has been extensively explored for various applications \cite{volakis,kolundvzija2002}. As a result, considerable research effort has been invested in understanding the nuances and ramifications of discretizing these equations. This includes various efforts to understand low frequency breakdown \cite{Wilton1981,Wu1995,Burton1995}, develop well conditioned formulations \cite{Taskinen2006}, introduce higher order basis sets \cite{graglia2015} and investigate accuracy and convergence \cite{warnick2008}, develop hierarchical basis \cite{peterson2006}, and so on. However, by and large, the problem has remained the same: how does one develop integral formulations that are well behaved across frequencies of interest especially when high discretization density is required to capture geometric features. To this end, several new SIE formulations and numerical techniques have been proposed; a partial listing of these includes the current and charge integral equation (CCIE) \cite{Taskinen2006}, augmented EFIE (A-EFIE) \cite{Qian2010},  Calder\'{o}n preconditioner \cite{Andriulli2008}, multi-resolution analysis \cite{Vipiana2005} and introducing loop-tree/star basis functions \cite{Wu1995,Burton1995,Vecchi1999,Zhao2000,Xiong2013}, and Debye sources \cite{Epstein2010,Epstein2013a,Epstein2015}. Recently, a decoupled potential integral equation (DPIE) based on Lorentz gauge has been proposed in \cite{Vico2014} that leads to a second-kind and stable formulation over a wide frequency band. DPIE was implemented numerically by using Nystr\"{o}m method in \cite{Vico2015}. A similar idea of exploiting generalized gauge based ${\bf A}$-$\Phi$ integral formulation was introduced in \cite{ChoChew2014} and its numerical implementation was presented in \cite{Liu2015}. 

The focus of this work is to build on an approach presented by Epstein and Greengard \cite{Epstein2010}. Their approach relies on using two scalar sources and building an SIE framework to solve for these unknowns. In their work, they demonstrated the efficacy of the approach when applied to a sphere and later to arbitrary shapes. The crux of this approach is to define two scalars, called Debye sources that are employed to represent Debye potentials which, in turn, can be generalized to represent currents on the arbitrary surface, not limited to the sphere surface. Two second-kind scalar integral equations can be derived by using this framework. However, implementing these equations in a discrete setting is challenging as one needs to find the inverse of the surface Laplacian or the Laplace-Beltrami operator that maps the unknown Debye sources to Debye potentials. A recent paper by Chernokozhin and Boag \cite{Chernokozhin2013} presented the first numerical discrete implementation of \cite{Epstein2010} on  piecewise smooth surfaces. In their work, the inverse of surface Laplacian operator was obtained by exploiting the finite difference scheme defined on structured grids on the piecewise flat surface. Besides, additional constraints have to be imposed in order to ensure the continuity of surface currents. The question we seek to ask is whether one can create a better/more accurate framework on surface representation that are described by an underlying local parameterization. 

To summarize the state of art of Debye sources based integral equation solvers, methods have been presented for analytical surface descriptions, and piecewise constant/lowest order Lagrangian surface description. The principal challenge is the lack of a robust surface Laplacian equation solver. The approach that we present overcomes this bottleneck by using a smooth definition of the surfaces. To this end, we take recourse to our recent work on isogeometric methods \cite{Li2016}. Iso-geometric/parametric analysis (IGA) has seen a resurgence in recent years, thanks in large part to Hughes \emph{et al.} \cite{Hughes2005}. The rationale for this research has been to provide a seamless interface between computer aided design (CAD) and computer aided engineering (CAE). In this framework, the same basis function set is employed to represent \emph{both} geometry and underlying physical quantities  residing on the geometry. While a bulk of the existing work has focused on using variations of splines, in what follows, we use subdivision surface representations \cite{Loop1987} that provide $C^2$ smoothness on the geometry almost everywhere. This representation was exploited to develop an IGA based EFIE solver \cite{Li2016} that relied on a surface Helmholtz decomposition, wherein the current is represented in terms of surface gradient and surface curl of potentials; henceforth referred to as P-EFIE.  In this paper, we will show that this additional smoothness provides the means to obtain a convergent inverse of the surface Laplacian \emph{without} imposing additional constraints to ensure continuity. 

The principal focus of this work is to exploit isogeometric basis sets to extend the idea of generalized Debye sources (GDS) \cite{Epstein2010,Epstein2013a,Epstein2015} to traditional EFIE.  This formulation will be referred to as GDS-EFIE. We will employ subdivision basis sets to create a discrete system to solve surface Laplacian equation embedded in GDS-EFIE first and then resolve the whole GDS-EFIE  system within the IGA framework. In this paper, we shall demonstrate 
\begin{itemize}
    \item performance of different numerical techniques for solving surface Laplacian equation,
    \item convergence of the eigenvalues estimation and the solution to  the inverse of Laplace-Beltrami operator,
    \item low frequency stability and better conditioning of the IGA GDS-EFIE solver,
    \item flexibility of the solution technique.
\end{itemize}
Parenthetically, we note that the proposed solver is straightforward to implement compared to the formulations proposed in \cite{Epstein2010,Chernokozhin2013}.

The remainder of this paper is organized as follows: Section II presents the classical EFIE briefly  and then proposes the variation of the EFIE formulation based on generalized Debye sources. Subdivision surfaces and functions are presented concisely in Section III. Numerical implementations of GDS-EFIE including surface Laplacian equation solvers are detailed in Section IV. Numerical examples validating the proposed approach are demonstrated in Section V. Finally, Section VI summarizes the contribution of this work as well as future directions.

\section{Formulations}
\subsection{Electric Field Integral Equation}
Consider a simply-connected PEC object $\Omega$ whose boundary $\partial\Omega$ is denoted by $\Gamma$ which is equipped with a unique outward pointing normal $\hat{n}$. 
This object is illuminated by incident fields (${{\bf{E}}^{inc}}$, ${{\bf{H}}^{inc}}$), and it results in scattered fields  exterior to $\Omega$. Total fields denoted by $({\bf E}, {\bf H})$ are composed of incident fields and scattered fields. The scattered fields due to the object can be obtained by using the equivalence principle, which posits an equivalent current ${\bf j} ({\bf r})$ for ${\bf r} \in \Gamma$ such that the total magnetic field just outside the surface $\Gamma^+$ satisfies the jump condition ${\hat n} \times {\bf H} = {\bf j}$, with a slight abuse of notation. The total electric field satisfies ${\hat n} \times {\bf E} =0$, which can be rewritten as 
\begin{equation}\label{eq:EFIE}
\hat{{n}}\times{\cal T}({{\bf{j}}}({\bf{r}}'))=-\hat{{n}}\times\hat{{n}} \times{{\bf{E}}^{inc}}({\bf{r}}),\qquad{\bf{r}}\in{\Gamma},
\end{equation}
that formulates the electric field integral equation (EFIE). In the above equation, the integral operator ${\cal{T}}$  is defined as 
\begin{equation}
{\cal T}({\bf{X}}({\bf{r}}))=\hat{n}\times ik \eta\int_{\Gamma} {\left[ {{\cal I} + \frac{{\nabla \nabla }}{{k^2}}} \right]g({\bf{r}},{\bf{r}}') \cdot{\bf{X}}({\bf{r}}')d{\bf{r}}'}
\end{equation}
in which $i$ is the imaginary unit ($i\equiv\sqrt{-1}$), ${k}$ is the wavenumber, ${\eta}$ is the intrinsic impedance of the background medium, ${\cal{I}}$ is the identity operator and
\begin{equation}
g({\bf{r}},{\bf{r}}')=\frac{{{e^{ik R}}}}{{4\pi{R}}}
\end{equation}
is the Green's function in free-space. In the above equation, ${R=|{\bf{r}} - {\bf{r}}'|}$ is the distance between the field point ${\bf{r}}$ and the source point ${\bf{r}}'$. An $e^{-i\omega t}$ time-dependence convention is assumed and suppressed throughout this paper. Here, $\omega$ is the angular frequency. The solution to the EFIE is typically effected by using method of moments (MoM) wherein surface current is represented by a set of vector basis functions, say the Rao-Wilton-Glisson basis functions \cite{Rao1982} which are equivalent to the lowest order Raviart-–Thomas functions \cite{Raviart1977}. Alternatives to this approach has been a topic of significant recent interest; these include using generalized method of moments (GMM) \cite{Nair2011,Dault2014}, subdivision surfaces \cite{Li2016}, discontinuous basis set \cite{Peng2013}
, and more recently, Debye sources \cite{Epstein2010}. All the aforementioned methods try to bring features into modeling electromagnetic scattering; but a common thread that ties GMM, MoM on subdivision surfaces, and Debye sources is the use of surface Helmholtz decomposition. In what follows, we prescribe  that generalized Debye sources can be exploited to solve the EFIE.  Ideas for using Debye sources were initiated by Epstein and others in a series of papers \cite{Epstein2013a,Epstein2010,Epstein2015}. While our approach is slightly different, genesis of ideas are rooted in the above citations. This approach necessitates the use of scalar representations for the sources. Next, we briefly describe EFIE based on generalized Debye sources.

\subsection{Generalized Debye Sources based EFIE}
Consider an arbitrary current ${\bf j}(\bf r)$ that exists on surface $\Gamma$. It is well known that this current may be written by using a surface Helmholtz decomposition as
\begin{equation}
\begin{split}
{\bf j}({\bf r})&=\nabla_{\Gamma}\Psi({\bf r})+\hat{n}\times\nabla_{\Gamma}\Phi({\bf r})+{\bf j}_H({\bf r})\\
&={\bf j}_1({\bf r})+{\bf j}_2({\bf r})+{\bf j}_H({\bf r}),
\end{split}
\end{equation}
where $\nabla_\Gamma$ is the surface gradient, $\Psi({\bf r})$ and $\Phi({\bf r})$ are scalar potentials and ${\bf j}_H({\bf r})$ is a harmonic vector field that  satisfies
\begin{equation}
\nabla_\Gamma\cdot{\bf j}_H({\bf r})=0,\quad
\nabla_\Gamma\cdot\left({\hat{n}\times\bf j}_H({\bf r})\right)=0.
\end{equation}
The harmonic component vanishes (${\bf j}_H({\bf r})\equiv 0$) on the surface of simply-connected geometry. The potentials can be related to two scalar sources  $p({\bf r})$ and $q({\bf r})$, the so-called generalized Debye sources via:
\begin{subequations}
\label{eq:SL1}
\begin{equation}
\nabla^2_{\Gamma}\Psi({\bf r})=\Delta_{\Gamma}\Psi({\bf r})=i\omega p({\bf r}),
\end{equation}
\begin{equation}
\nabla^2_{\Gamma}\Phi({\bf r})=\Delta_{\Gamma}\Phi({\bf r})=-i\omega q({\bf r}).
\end{equation}
\end{subequations}
It is noted that $p({\bf r})$ has direct physical meaning here, i.e., surface charge density unlike the one in \cite{Epstein2010,Chernokozhin2013}, and 
$\Delta_{\Gamma}$ is the surface Laplacian or Laplace-Beltrami operator. Then the currents can be rewritten as
\begin{equation}
{\bf j}_1({\bf r})=i\omega\nabla_{\Gamma}\Delta_{\Gamma}^{-1}p({\bf r}),\quad
{\bf j}_2({\bf r})=-i\omega\hat{n}\times\nabla_{\Gamma}\Delta_{\Gamma}^{-1}q({\bf r}),
\end{equation}
in which $\Delta^{-1}_{\Gamma}$ is the inverse of Laplace-Beltrami operator. As surface current is represented by generalized Debye sources, EFIE (\ref{eq:EFIE}) can be rewritten in terms of $p({\bf r}')$ and $q({\bf r}')$ as follow: 
\begin{equation}
\hat{n}\times{\cal T}\left ({\mathcal{C}}({p}({\bf{r}}'))-\hat{{n}}'\times{\mathcal{C}}({q}({\bf r}'))\right)=-\hat{n}\times\hat{n}\times{{\bf{E}}^{inc}}({\bf{r}}),\,{\bf{r}}\in{\Gamma},
\label{eq:C-EFIE}
\end{equation}
where
\begin{equation}
\mathcal{C}(f({\bf r}))=i\omega\nabla_{\Gamma}\Delta_{\Gamma}^{-1}f({\bf r}).
\end{equation}
Equation (\ref{eq:C-EFIE}) can be considered as generalized Debye sources based EFIE (GDS-EFIE). It is noted that there are always non-trivial solutions to equation (\ref{eq:SL1}),  $\Psi$=constant and $\Phi$=constant. To ensure uniqueness of the inverse of Laplace-Beltrami operator, 
$\mathcal{M}_\Gamma$, a space of mean-zero function on $\Gamma$ can be defined first:
\begin{equation}
\mathcal{M}_\Gamma=\left\lbrace f:\Gamma\to \mathbb{R}|\int_\Gamma f({\bf r})d{\bf r}=0\right\rbrace.
\end{equation}
Then surface Laplacian operator $\Delta_\Gamma$ might be invertible if a map from $\mathcal{M}_\Gamma$ to itself on $\Gamma$ can be established \cite{Epstein2010}. Given that function space of generalized Debye sources belongs to $\mathcal{M}_\Gamma$, a unique solution to the inverse of $\Delta_\Gamma$ can be achieved by solving the following boundary value problem:
\begin{subequations}
\label{eq:SL}
\begin{equation}
\nabla_{\Gamma}^2\Psi({\bf r})=f({\bf r}),
\label{eq:SL2}
\end{equation}
\begin{equation}
\label{eq:con}
\int_{\Gamma}\Psi({\bf r}) d{\bf r}=0.
\end{equation}
\end{subequations}
Equation (\ref{eq:con}) can be considered as a constraint to surface Laplacian equation (\ref{eq:SL2}) in order to obtain a unique solution. Similar comments are valid for $\Phi({\bf r})$, and are omitted in the above discussion for the sake of brevity. 

It is worth noting that both discretization and seeking the inverse of the surface Laplacian operator are non-trivial. First, one needs to find a smooth enough representation for each scalar quantity on complex surfaces so that the resulting currents satisfy the continuity condition. The work in \cite{Chernokozhin2013} introduced additional conditions to guarantee continuity. Our proposed method, however, relies on a smooth basis set ($C^2$ almost everywhere and $C^1$ at isolated points) for scalars with the help of the subdivision surface representation. Second, the direct inverse of surface Laplacian operator doesn't exist if no constraint is imposed to remove the known constant-value null space. In the next section, subdivision surface and subdivision basis set are reviewed briefly.

\section{Subdivision Surfaces and Functions}
As a shape description, subdivision surfaces technique has been explored extensively in computer graphics, especially the animation industry. Even though non-uniform rational B-splines (NURBS) is built in most CAD systems as an industry standard, NURBS results in a smooth description in the interior of a patch but {\em only} $C^0$ or even worse across the boundary between patches \cite{Bazilevs2010}. In comparison, the limit surface generated by subdivision schemes is $C^2$ almost everywhere except at finite points (irregular vertices) where the description is $C^1$. Thus, one can exploit subdivision basis to resolve GDS-EFIE within IGA framework without imposing any additional conditions to ensure continuity.

\begin{figure}[t]
\centering
\includegraphics[width=1.8in]{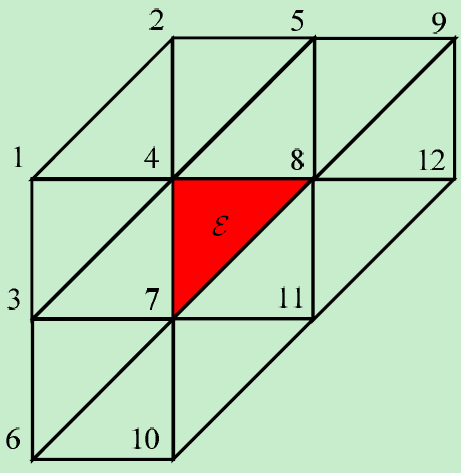}
\caption{A regular triangular patch defined by vertices in its 1-ring vertices.}
\label{fig:reg_tri}
\end{figure}

Since triangular tessellations are omnipresent in SIE solvers for EM problems, we employ function spaces defined by the Loop subdivision scheme \cite{Loop1987}. Consider an initial primal mesh denoted by $P_0$ at level 0 which consists of a set of vertices and connectivity map. The valence of a given vertex is defined as the number of triangles incident on itself. A 1-ring of a vertex consists of all vertices of these triangles. A vertex is considered as a regular one if its valence equals to 6. Otherwise, it is called an \emph{irregular} or \emph{extraordinary} vertex. A triangle is regular if its vertices are  \emph{all} regular, and irregular otherwise. A limit surface can be generated by recursive refinements of primal mesh $P_0$. Specifically, there are $\mathcal{N}_t$ triangles in $P_0$ and Loop subdivision scheme generates triangular tessellations of level $k$ recursively including $4^k\mathcal{N}_t$ new triangular patches by inserting new vertices at the edge midpoints (in parameter domain sense) and subdividing a triangle of level $k-1$ into four sub-triangles of level $k$. It is noted that every newly inserted vertices are regular and only irregular vertices at level 0 still remain irregular. After each subdivision, the position of {\emph{every}} vertex will be recomputed and each new triangle patch can be parameterized by 1-ring of the patch(union of vertices of incident triangles to its three vertices).  Fig. \ref{fig:reg_tri} illustrates a regular triangle $\mathcal{E}$ defined by its 1-ring vertices indexed from 1 to 12. 

When a nodal value function $\xi_n({\bf r})$ is associated with $n$th vertex, a regular triangular patch can be evaluated as 
\begin{equation}
{\bf S}({\bf r}(u,v))=\sum\limits_{n=1}^{12}{\bf t}_n{\xi_n}({\bf r}(u,v))
\end{equation}
where $(u,v)$ denotes a  pairwise coordinate on a parameterization chart and ${\bf t}_n\in\mathbb{R}^3$ is the position vector of $n$th vertex. For a regular triangular patch, $\xi_n({\bf r})$ is a quartic box spline tabulated in \cite{Stam1998}. Since a triangular patch is defined by its 1-ring vertices, the scalar function $\xi_n({\bf r})$ can affect on 2-ring domain of vertex ${\bf t}_n$ (union of the 1-rings of the 1-ring). It is zero outside the vertex's 2-ring. Thus, $\xi_n({\bf r})$ has a compact support and $\xi_n({\bf r})\in C^2_0$. For an irregular vertex, $\xi_n({\bf r})$ is still a piecewise polynomial function, however, it has a infinite number of segments towards the irregular vertex itself. As such, $\xi_n({\bf r})$ will degrade to $C^1$ smoothness only at the irregular vertex. For more details on evaluating $\xi_n({\bf r})$ associated with an irregular vertex, we refer to the reference \cite{Stam1998,Li2016}.

\section{Algorithmic Flowchart}
In this section, we will describe the procedure that we will exploit to solve GDS-EFIE \eqref{eq:C-EFIE}. The solution to this equation proceeds from right to left. Specifically, the unknowns in the system are the scalar sources, $p({\bf r})$ and $q({\bf r})$. These are then mapped to potentials, $\Psi ({\bf r})$ and $\Phi ({\bf r})$, which are in turn mapped to the equivalent electric current ${\bf j}({\bf r})$. Finally, the integral operator ${\cal T}$ maps the current onto the scattered electric field. In what follows, each of these mapping operations is discussed in sequence, starting with the inverse of the surface Laplacian and followed by representation of current and then the discretization of the operator. 

To start,  we  assume that  given $\xi_n ({\bf r})$ for $n$th vertex of the primal mesh $M_0$ with $N$ vertices in total, the unknown generalized Debye sources $p({\bf r})$ and $q({\bf r})$ in GDS-EFIE can be represented by 
\begin{equation}
p({\bf r})\approx\sum\limits_{n=1}^{N}a_{1,n}\xi_n({\bf r}),\qquad
q({\bf r})\approx\sum\limits_{n=1}^{N}a_{2,n}\xi_n({\bf r}),
\end{equation}
where $a_{1,n}$ and $a_{2,n}$ are unknown coefficients. 
Since surface current representations in terms of generalized Debye sources require the inverse of Laplace-Beltrami operator, three different numerical techniques for solving surface Laplacian equation  are discussed next. In what follows, we only illustrate the solution for $\Psi ({\bf r})$; the solution for $\Phi ({\bf r})$ can be obtained in a similar manner. 

\subsection{Surface Laplacian Solvers}
It can be shown that the solution of equation (\ref{eq:SL2}) will minimize the functional
\begin{equation}
F(\Psi)=\frac{1}{2}\int_\Gamma\nabla_\Gamma^2\Psi({\bf r})d{\bf r}-\int_\Gamma f({\bf r})\Psi({\bf r})d{\bf r}
\end{equation}
which can be achieved by imposing the stationary requirement ${\delta}F=0$. To satisfy this requirement, we start with representing the potential $\Psi({\bf r})$ by using subdivision basis functions such that $\Psi({\bf r})\approx\sum\nolimits_{n=1}^{N}b_{1,n}\xi_n({\bf r})$. Then a $N\times N$ matrix equation corresponding to  the weak form of variational problem $\delta F=0$ can be written as:
\begin{equation}\label{eq:Gram}
{\bf G}{\bf b}_1={\bf h}
\end{equation}
where 
\begin{subequations}
\begin{equation}\label{eq:GramMat}
{G}_{m,n}=\int_{\Gamma_m\cap\Gamma_n}{\nabla_s\xi_m({\bf r})\cdot\nabla_s\xi_n({\bf r})}d{\bf r}
\end{equation}
and
\begin{equation}
\quad{h_n}=-\int_{\Gamma_n}{\xi_n({\bf r})f({\bf r})d{\bf r}}.
\end{equation}
\end{subequations}
Here, $\Gamma_{m(n)}$ denotes the support of basis function $\xi_{m(n)}$. 
Thus far, only discretization of equation (\ref{eq:SL2}) is considered. Since matrix ${\bf G}$ is rank deficient by one due to the existing of one-dimensional null space, one needs to solve the equation (\ref{eq:SL2}) in tandem with (\ref{eq:con}). In this work, three different techniques including least squares (LSQ), penalty and Lagrange multiplier (LM) are studied. Comparisons in effectiveness and efficiency are also made among these methods in the results section. 

\subsubsection{Least Squares Solution}
Substituting the representation of the potential function to the constraint (\ref{eq:con}) and letting $\int_{\Gamma_n}{\xi_n}({\bf r})d{\bf r}=c_n$, we have  
\begin{equation}
\sum\limits_{i=1}^{N}b_{1,n} c_n=0.
\end{equation}
Therefore, the last coefficient $b_{1,N}$ can be expressed in a linear combination of others as  $b_{1,N}=- {{\bf b}'_1}^T {\bf c}' /c_N$, where ${\bf b}'_1=(b_{1,1},b_{1,2},...,b_{1,N-1})^T$ and ${\bf c}'=(c_1,c_2,...,c_{N-1})^T$. 
As a result, one can define a sparse transformation matrix
\begin{equation}
{\bf T}=
 \begin{bmatrix}
 1 & 0 & \cdots & 0 \\
 0 & 1 & \cdots & 0 \\
 \vdots & \vdots & \ddots & \vdots \\
 0 & 0 & \cdots & 1 \\
 -c_1/c_N & -c_2/c_N & \cdots & -c_{N-1}/c_N
 \end{bmatrix}
\end{equation}
that maps ${\bf b}'_1$ onto ${\bf b}_{1}$ via ${\bf T}{\bf b}'_1={\bf b}_1$. The constraint (\ref{eq:con}) together with the discrete system can be  rewritten as
\begin{equation}
{\bf G}{\bf T}{\bf b}'_1={\bf h}.
\end{equation}
This is an overdetermined $N\times(N-1)$ matrix equation with respect to ${\bf b}'_1$ and may be solved approximately via least squares method as follow:
\begin{equation}\label{eq:LSQ}
{\bf T}^{T}{\bf G}{\bf T}{\bf b}'_1={\bf T}^T{\bf h}.
\end{equation}
It is noted that the product of ${\bf T}^{T}{\bf G}{\bf T}$ is not necessarily sparse. However, as each involved matrix of the product is sparse, one may obtain an iterative solution with ${\cal O} (N)$ for per matrix vector multiplication (MVM). 

\subsubsection{Penalty Method}
While the above approach takes the constraint (\ref{eq:con}) directly into account, an alternate approach is augmenting the functional with a penalty constraint. The functional defined earlier may be modified to include a penalty function as follows: 
\begin{equation}
P(\Psi)=F(\Psi)+\frac{\gamma}{2}\left(\int_{\Gamma}\Psi({\bf r})d{\bf r}\right)^2
\end{equation}
where $\gamma$ is a prescribed parameter. 
Hence, the matrix equation (\ref{eq:Gram}) might be modified as 
\begin{equation}\label{eq:Pen}
{\bf G}^p{\bf b}_1={\bf h}
\end{equation}
where
\begin{equation}
\begin{split}
G_{m,n}^p=&\int_{\Gamma_m\cap\Gamma_n}{\nabla_s\xi_m({\bf r})\cdot\nabla_s\xi_n({\bf r})}d{\bf r}\\&+\gamma\int_{\Gamma_m}\xi_m({\bf r})d{\bf r}\int_{\Gamma_n}\xi_n({\bf r})d{\bf r}.
\end{split}
\end{equation}
Here, $p$, the subscript of $G_{m,n}^p$, indicates the matrix entry of penalty method. Similar to LSQ method, the final system is a full matrix. 
\subsubsection{Lagrange Multiplier Method}
Another viable alternative to the above is to employ a Lagrange multiplier; as before, this involves a minor modification to the functional that is to be minimized, viz., 
\begin{equation}
L(\Psi,\beta)=F(\Psi)+\beta\int_{\Gamma}\Psi({\bf r})d{\bf r}.
\end{equation}
Similarly, the stationary requirement $\delta L=0$ results in a unique ${\bf b}_1$ and $\beta_0$ that will minimize the above functional. Letting $\partial L/\partial b_{1,n}=0$ and $\partial L/\partial\beta=0$, one can obtain a matrix equation with an augmented dimension of $(N+1)\times(N+1)$:
\begin{equation}\label{eq:LM}
 \begin{bmatrix}
{\bf G} & {\bf c} \\
{\bf c}^T & 0 \\
 \end{bmatrix}
 \begin{bmatrix}
{\bf b}_1 \\
{\beta} \\
 \end{bmatrix}
 = \begin{bmatrix}
{\bf h} \\
{0} \\
 \end{bmatrix}.
\end{equation}
It is noted that the resulting matrix of Lagrange multiplier method is sparse when $N\gg 1$. Hence, the complexity for both memory and MVM scales as $\mathcal{O}(N)$.

\subsection{Discretization of GDS-EFIE}
Thus far, we have prescribed three different approaches for seeking the inverse of the surface Laplacian operator based on the following representation:
\begin{equation}
\label{eq:ExpanPsi}
\Psi({\bf r})\approx\sum\limits_{n=1}^{N}b_{1,n}\xi_n({\bf r}),\qquad
\Phi({\bf r})\approx\sum\limits_{n=1}^{N}b_{2,n}\xi_n({\bf r}).
\end{equation}
It follows that once the inverse is found, we can obtain the coefficients $b_{1,n}$ and $b_{2,n}$ which are related to $a_{1,n}$ and $a_{2,n}$ that are involved to represent the unknown Debye sources $p ({\bf r})$ and $q ({\bf r})$. Then, surface currents ${\bf j}_1({\bf r})$ and ${\bf j}_2({\bf r})$ can be expressed as
\begin{subequations}
\begin{equation}
{\bf j}_1({\bf r})\approx\sum\limits_{n=1}^{N}a_{1,n}{\bf j}_{1,n}({\bf r})=  i\omega\sum\limits_{n=1}^{N}b_{1,n}\nabla_{\Gamma}\xi_n({\bf r}),
\end{equation}
\begin{equation}
{\bf j}_2({\bf r})\approx\sum\limits_{n=1}^{N}a_{2,n}{\bf j}_{2,n}({\bf r}) = -i\omega\sum\limits_{n=1}^{N}b_{2,n}\hat{n}\times\nabla_{\Gamma}\xi_n({\bf r}).
\end{equation}
\end{subequations}
Using these expressions for the current and testing GDS-EFIE \eqref{eq:C-EFIE}
with $\nabla_\Gamma\xi_n({\bf r})$ and $\hat{n}\times\nabla_\Gamma\xi_n({\bf r})$ results in a system that reads as \cite{Li2016} 
\begin{equation}\label{eq:impMatSys}
\begin{bmatrix}
{\bf Z}^{11} & {\bf Z}^{12} \\
{\bf Z}^{21} & {\bf Z}^{22}  \\
\end{bmatrix}
\begin{bmatrix}
{\bf b}_{1}\\
{\bf b}_{2}\\
\end{bmatrix}
=\begin{bmatrix}
{\bf e}_{1}\\
{\bf e}_{2}\\
\end{bmatrix}
\end{equation}
where
\begin{subequations}
\begin{equation}
\begin{split}
&Z^{st}_{mn}=-i\omega\mu\int_{\Gamma_m}{\bf j}_{s,m}({\bf r})d{\bf r}\cdot\int_{\Gamma_n}g({\bf r},{\bf r}'){\bf j}_{t,n}({\bf r}')d{\bf r}'\\
&+\frac{i\delta_{s1}\delta_{t1}}{\omega\epsilon}\int_{\Gamma_m}\nabla_{\Gamma}\cdot{\bf j}_{s,m}({\bf r})d{\bf r}\cdot\int_{\Gamma_n}g({\bf r},{\bf r}')\nabla_{\Gamma}\cdot{\bf j}_{t,n}({\bf r}')d{\bf r}'
\end{split}
\end{equation}
\begin{equation}
e_{s,n}=\int_{\Gamma_n}{\bf j}_{s,n}({\bf r})\cdot{\bf E}^{inc}({\bf r})d{\bf r}
\end{equation}
and
\begin{equation}
\delta_{ij} = \begin{cases}
  1, & \mbox{if } i=j \\
  0,  & \mbox{if } i\neq j.
\end{cases}
\end{equation}
\end{subequations} 
Then an iterative solution for the unknown coefficients $a_{i,n}$ for $i=1,2$ proceeds via two sets of iterations (i) inverse of the surface Laplacian to map from $a_{i,n} \longrightarrow b_{i,n}$ for $i=1,2$, and (ii) convergence of \eqref{eq:impMatSys}. Together, this results in solutions to $a_{i,n}$. Next, we discuss a series of results that address convergence of surface Laplacian solvers as well as solutions to the proposed EFIE. 

\section{Numerical examples}
\subsection{Performance Comparison of Surface Laplacian Solvers }
To test the performance of three aforementioned surface Laplacian solvers, three metrics are employed: (a) the residual of left hand side of constraint (\ref{eq:con}) which is supposed to be zero exactly, (b) the condition number of the resulting matrix system and (c) the convergence rate for each solver. For a given right hand side of (\ref{eq:SL}) $f\in\mathcal{M}_\Gamma$, matrix equations (\ref{eq:LSQ}), (\ref{eq:Pen}) and (\ref{eq:LM}) are solved iteratively with maximum of iteration count set to 600 and tolerance $10^{-13}$, respectively. Bi-conjugate gradients (BiCG) method is employed as an iterative solver for surface Laplacian equation. In this paper, condition number of matrix ${\bf A}$ is defined in matrix 2-norm sense, namely,
\begin{equation}
\kappa({\bf A})=||{\bf A}||_2||{\bf A}^{-1}||_2.
\end{equation}
For penalty method, the prescribed parameter $\gamma$ ranges from 1 to $10^9$ and is sampled at every order. Finally, the examples chosen are akin to those used in computer graphics \cite{Juttler2016}. 

\begin{figure}[t]
\centering
\includegraphics[width=3.5in]{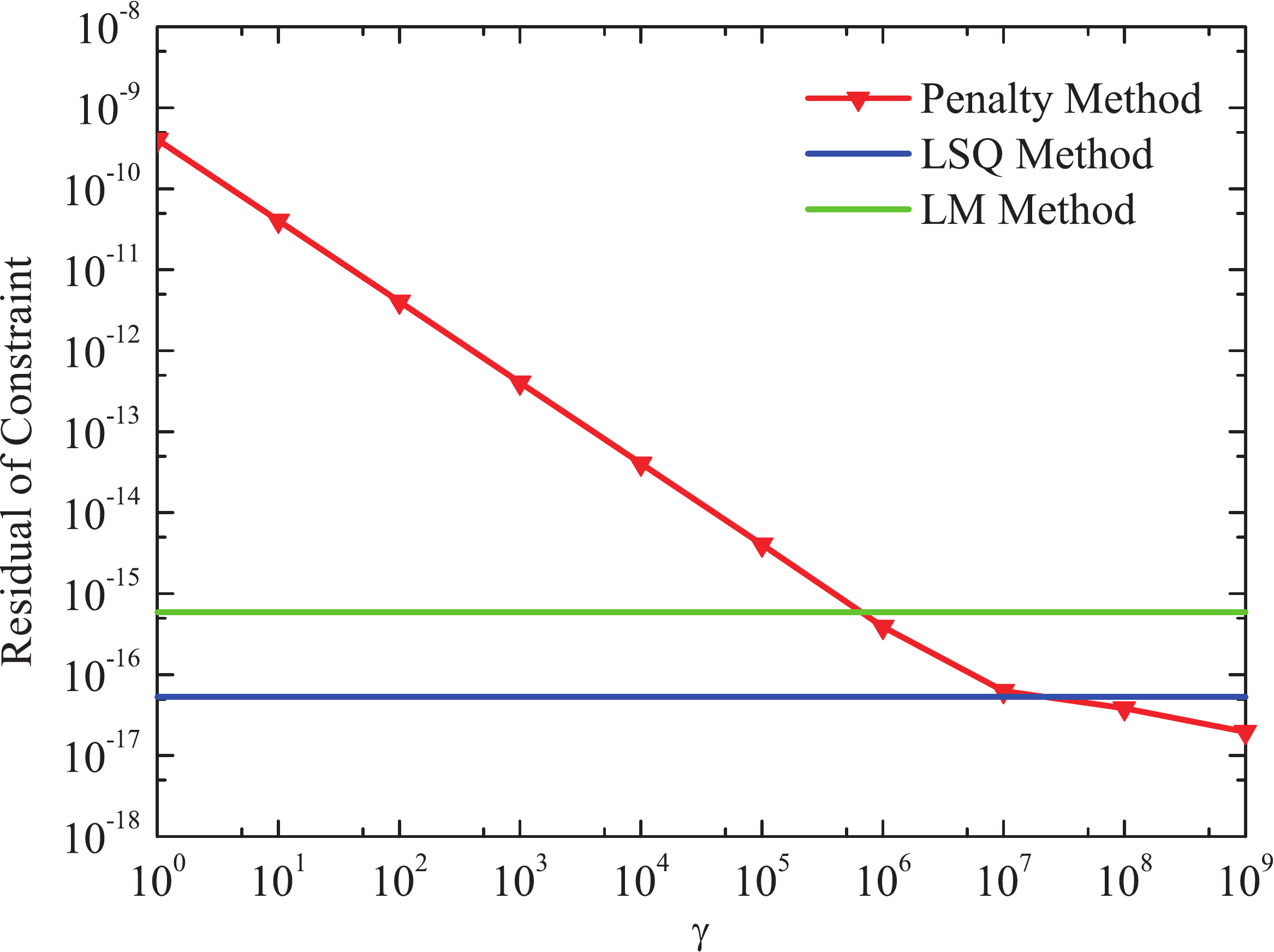}
\caption{Residual of left hand side of constraint (\ref{eq:con}) of different surface Laplacian solvers for a sphere.}
\label{fig:Res_sph}
\end{figure}

\begin{figure}[t]
\centering
\includegraphics[width=3.5in]{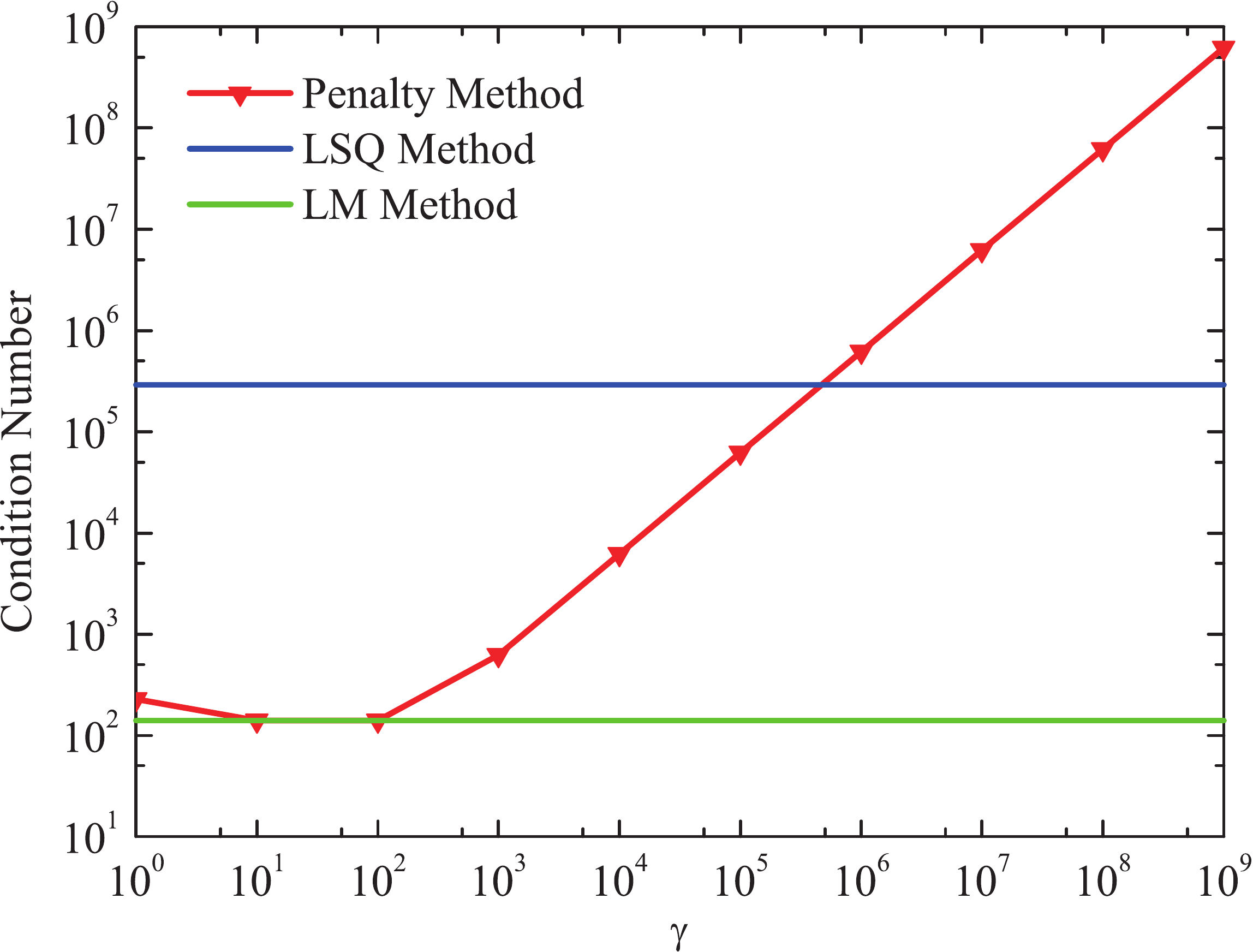}
\caption{Condition number of resulting matrix system of different surface Laplacian solvers for a sphere.}
\label{fig:CN_sph}
\end{figure}

\begin{figure}[t]
\centering
\includegraphics[width=3.5in]{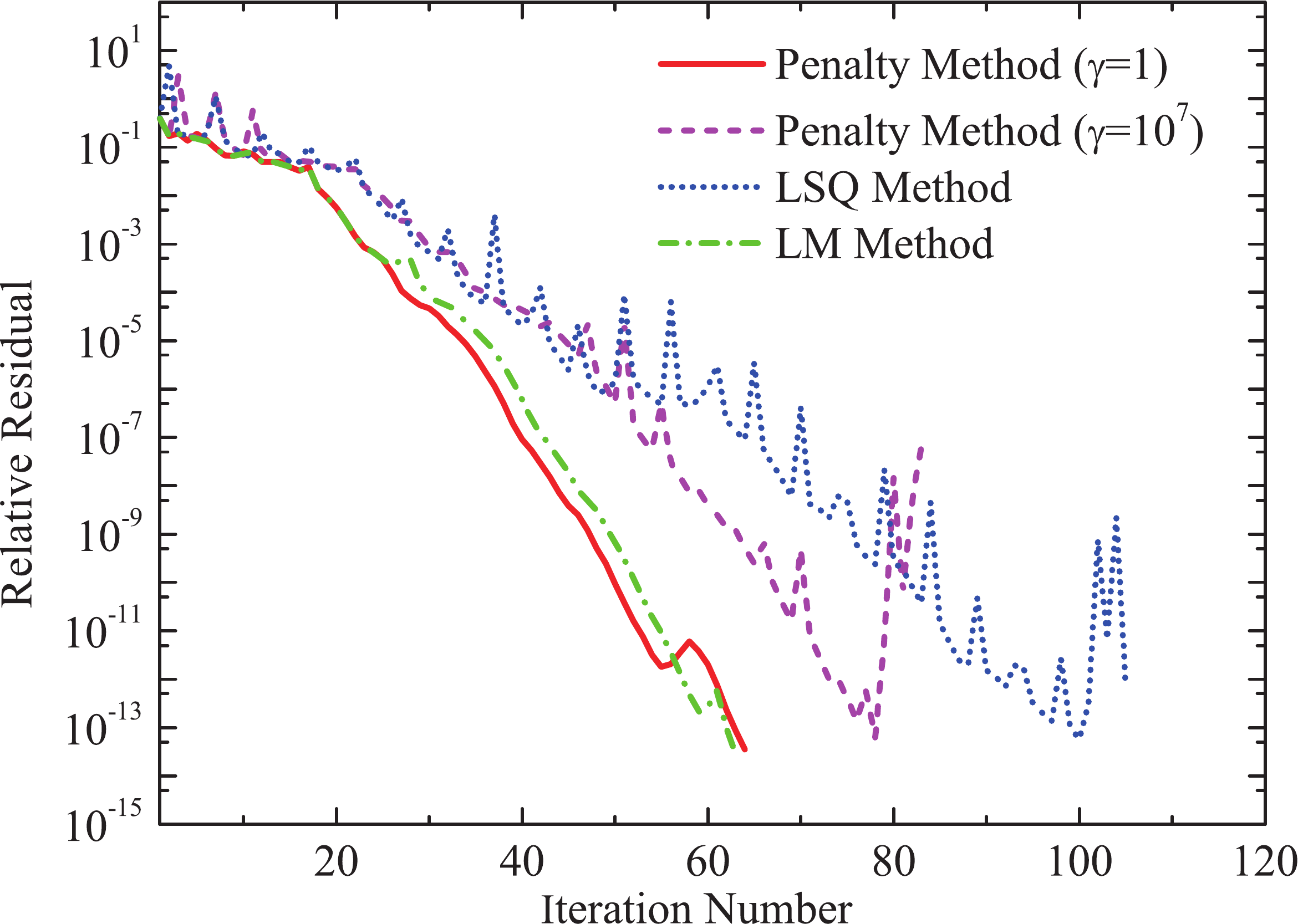}
\caption{Convergence history of different surface Laplacian solvers for a sphere.}
\label{fig:Iter_sph}
\end{figure}

A sphere with radius $r=1$m is first analyzed. The sphere surface is discretized into 5120 triangular elements.  Hence,  there are 2562 vertices. The residual of constraint of penalty method as a function of parameter $\gamma$ is depicted in Fig. \ref{fig:Res_sph}. For better comparison and illustration, the residuals of least square (LSQ) and Lagrange multiplier (LM) methods are also plotted in the figure even though they are both independent of $\gamma$. As observed, the solution of penalty method is more accurate as $\gamma$ increases. It is also evident that the accuracy of both LSQ and LM method is very high.
The condition number of the matrix system of three different solvers is plotted in Fig. \ref{fig:CN_sph}, respectively. From the figure, it is evident that the condition number of penalty method increases rapidly as $\gamma$ increases. The condition number of LM method is the lowest among three solvers.  
Fig. \ref{fig:Iter_sph} depicts the relative residuals of three solvers as iteration number increases. For penalty method, two curves corresponding to $\gamma=1$ and $\gamma=10^7$ respectively are plotted. 

\begin{figure}[t]
\centering
\includegraphics[width=3.5in]{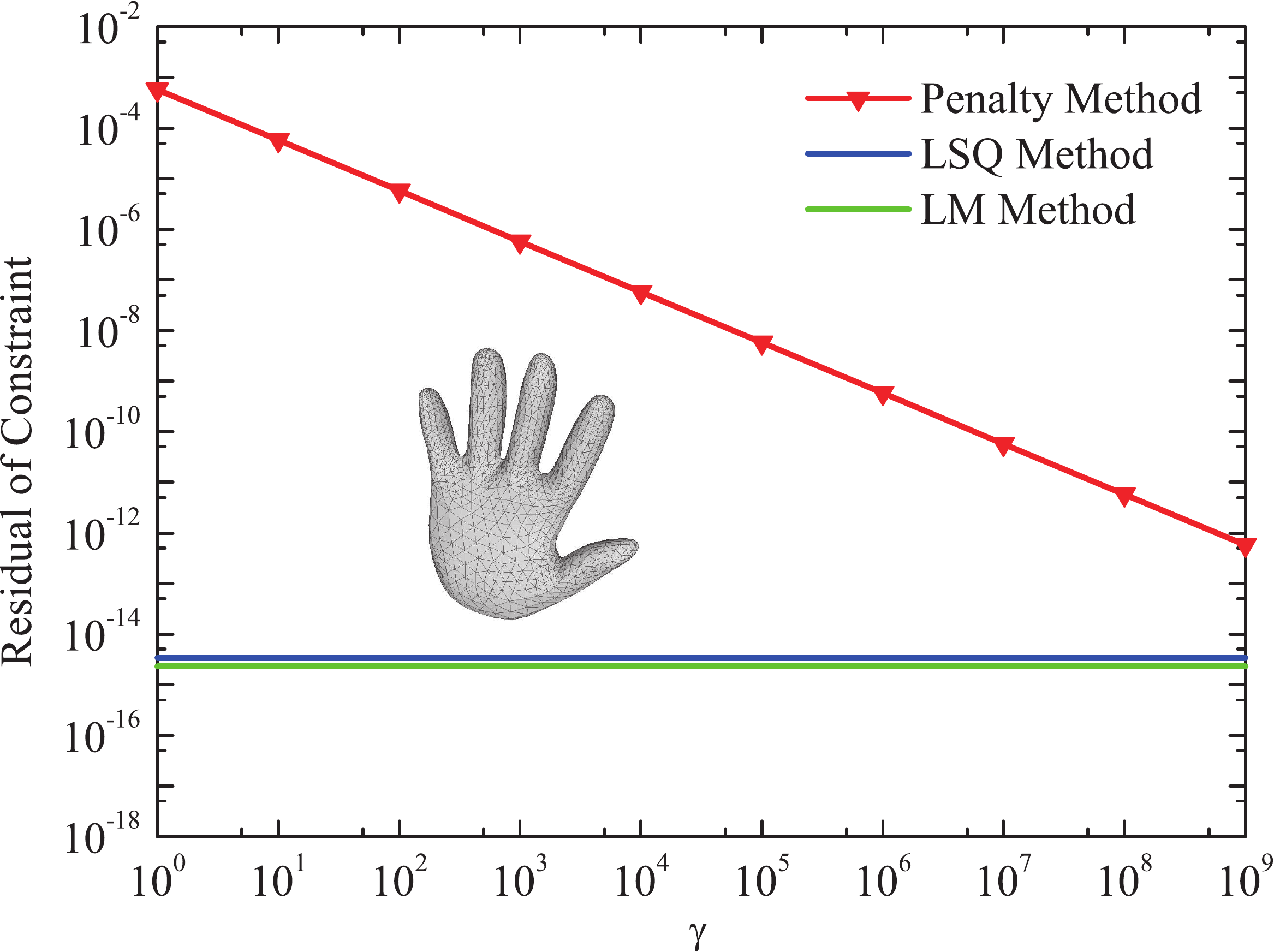}
\caption{Residual of left hand side of constraint (\ref{eq:con}) of different surface Laplacian solvers for a hand shape object. Inset: A meshed hand shape object.}
\label{fig:Hand_Res}
\end{figure}

\begin{figure}[t]
\centering
\includegraphics[width=3.5in]{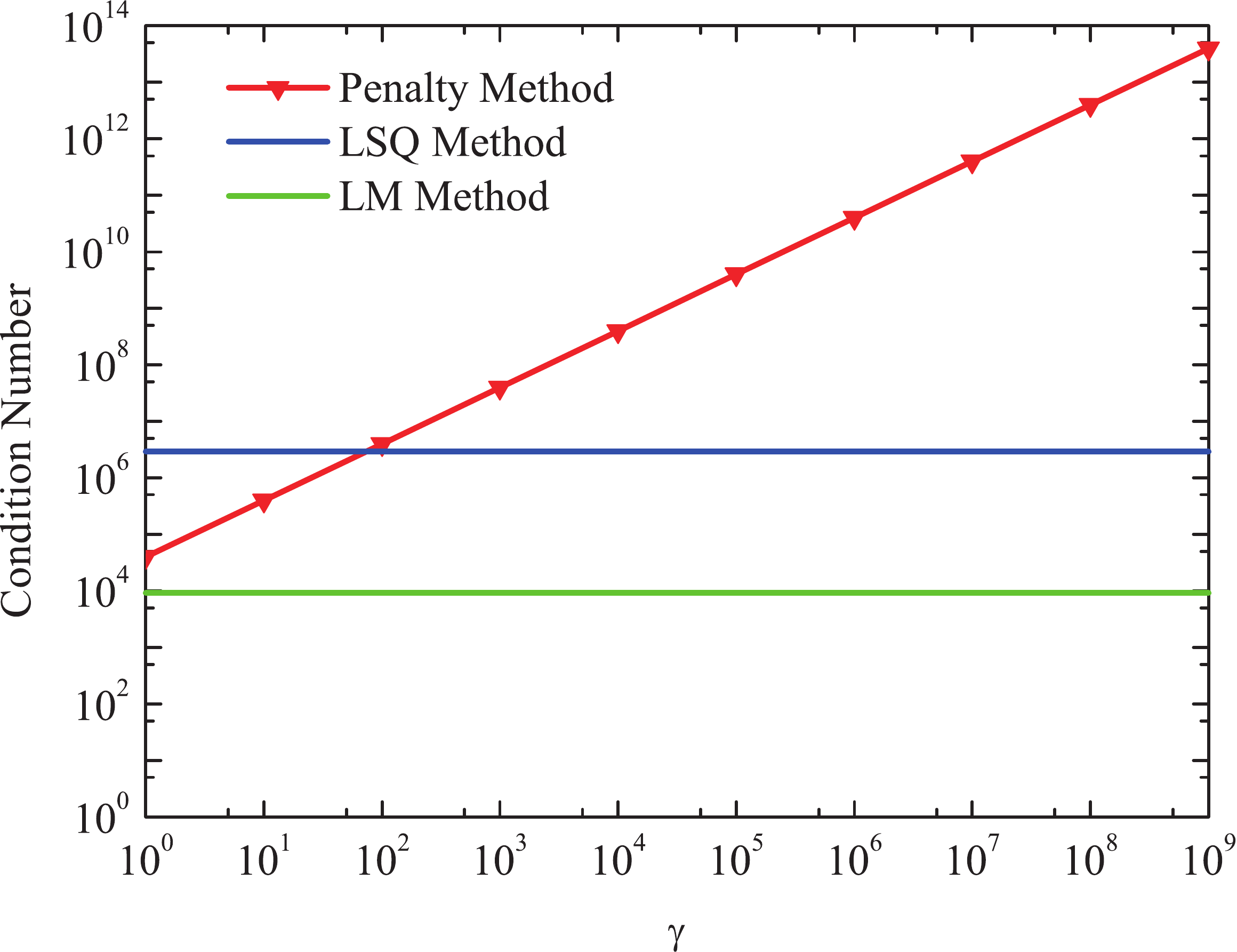}
\caption{Condition number of resulting matrix system of different surface Laplacian solvers for a hand shape object.}
\label{fig:CN_Hand}
\end{figure}

\begin{figure}[t]
\centering
\includegraphics[width=3.5in]{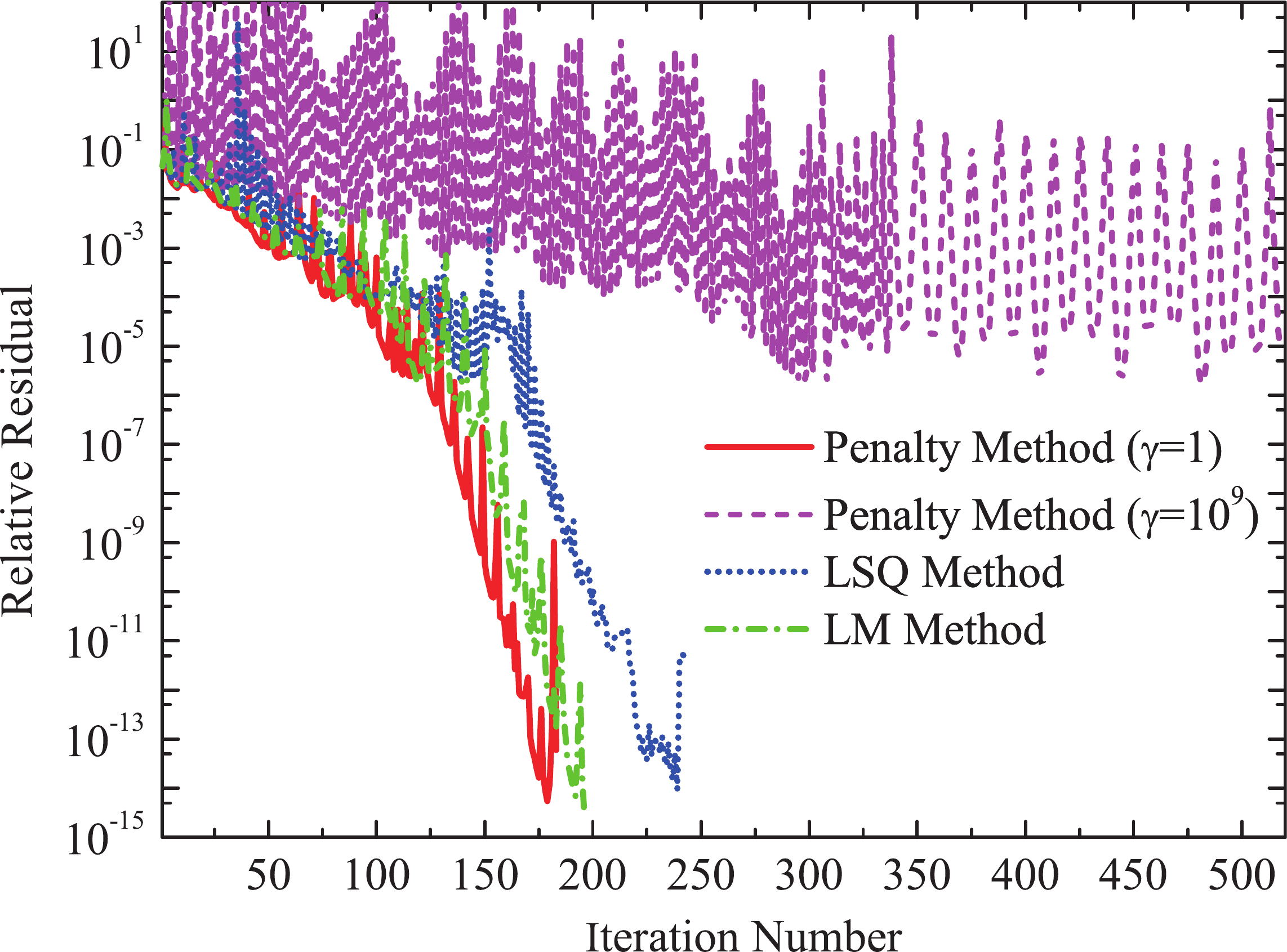}
\caption{Convergence history of different surface Laplacian solvers for a hand shape object.}
\label{fig:Iter_Hand}
\end{figure}

In the second example, we consider a hand shape object meshed as shown in inset of Fig. \ref{fig:Hand_Res}. The hand is discretized by 2880 triangular elements and 1442 vertices. Fig. \ref{fig:Hand_Res} demonstrates the residuals of constraint (\ref{eq:con}) and it is evident that the accuracy of penalty method is the worst. Even when $\gamma=10^9$, the residual of penalty method is two orders larger than LSQ and LM method. Again, as parameter $\gamma$ increases, the condition number of penalty method will be larger as shown in Fig. \ref{fig:CN_Hand}. LM method outperforms once more. Finally, Fig. \ref{fig:Iter_Hand} depicts the residual changes as iteration number increases for three solvers. It is noted that the accuracy of penalty method is poor when $\gamma=1$ although it can achieve the tolerance fast. Hence, it is not straightforward to provide a guideline how to choose parameter $\gamma$ for penalty method. The accuracy and matrix conditioning will go to the opposite side when $\gamma$ increases.  

Taking the three metrics into account, we posit that the performance of LM method is optimal amongst the three solvers. Moreover, the implementation of LM is not expensive since the complexity of per MVM only scales as $\mathcal{O}(N)$.

\subsection{Convergence Test of the Surface Laplacian}

\begin{figure}[t]
\centering
\includegraphics[width=3.5in]{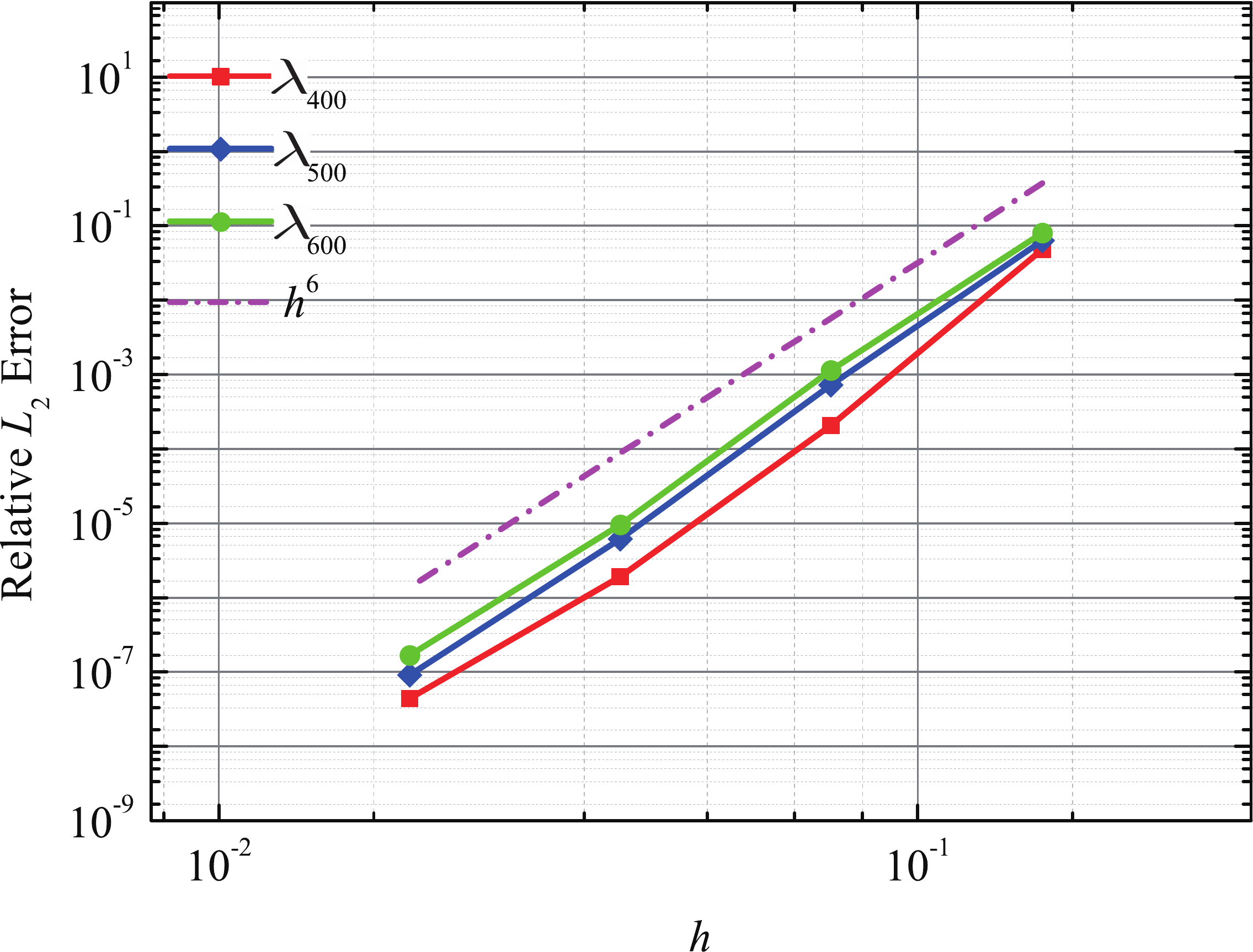}
\caption{Convergence for generalized eigenvalue on a sphere.}
\label{fig:Eig_Sph_LM}
\end{figure}

\begin{figure}[t]
\centering
\includegraphics[width=3.5in]{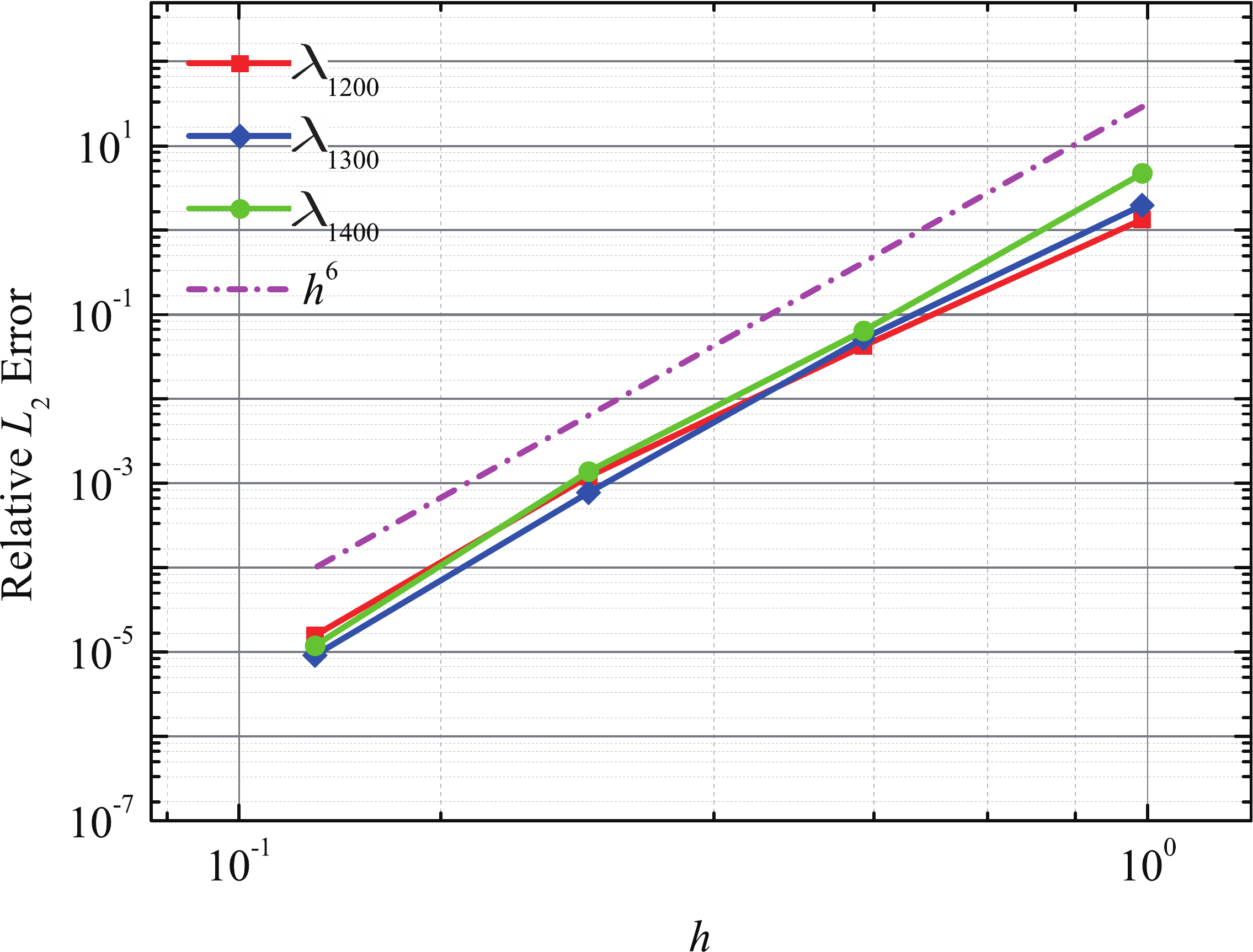}
\caption{Convergence for generalized eigenvalue on a hand shape object.}
\label{fig:Eig_Hand_LM}
\end{figure}

\begin{figure}[t]
\centering
\includegraphics[width=3.5in]{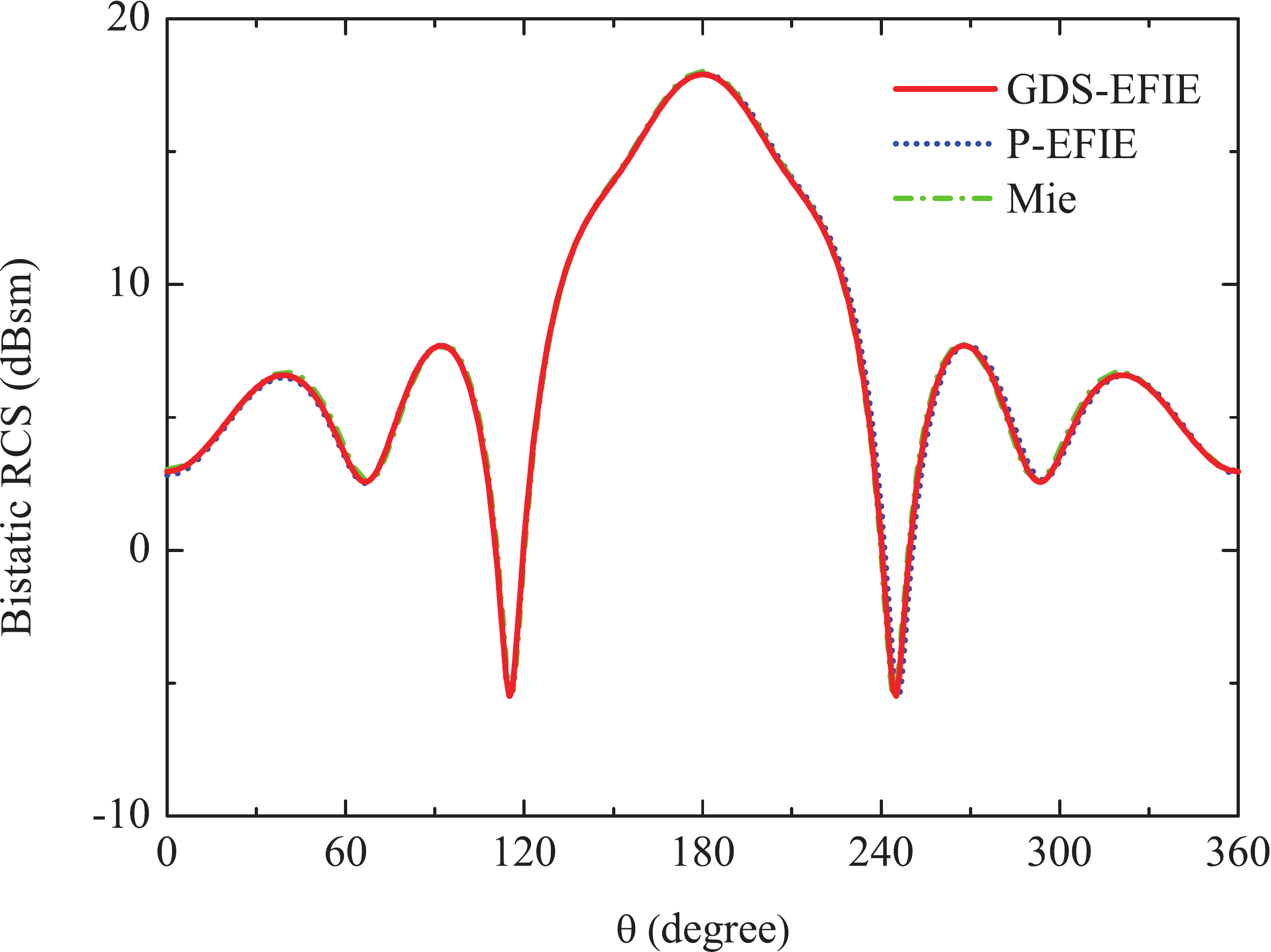}
\caption{Bistatic RCS solutions at $\phi=0$ cut for a sphere with radius $r=0.67\lambda$.}
\label{fig:Sph_RCS}
\end{figure}

While the above results presented how well different methods worked, it is also important that the results by themselves converge with spatial refinement. To this end, we formulate a generalized eigenvalue problem and exploit this as a metric to study convergence, specifically that of the LM method. The generalized eigenvalue can be formulated as 
\begin{equation}
{\bf P}{\bf v}=\lambda_n{\bf Q}{\bf v}
\label{eq:Eig_LM}
\end{equation}
where $\lambda_n$ is $n$th generalized eigenvalue to solve, ${\bf v}$ is the corresponding eigenvector and
\begin{equation}
{\bf P}=
 \begin{bmatrix}
{\bf G} & {\bf c} \\
{\bf c}^T & 0 \\
 \end{bmatrix},\quad
{\bf Q}=
 \begin{bmatrix}
{\bf M} & {\bf 0} \\
{\bf 0} & 1 \\
 \end{bmatrix}.
\end{equation}
 Here, ${\bf M}$ is a mass matrix with its entry defined as
\begin{equation}
M_{m,n}=\int_{\Gamma_m\cap\Gamma_n}{\xi_m({\bf r})\cdot\xi_n({\bf r})}d{\bf r}.
\end{equation}
We solve the generalized eigenvalue problems for the both sphere and hand shape object analyzed earlier. To study the experimental order of convergence, an initial control mesh with 642 vertices for the sphere is used and three subsequent subdivisions are conducted which can generate 2562, 10242 and 40962 vertices, respectively. The reference solution comes from one more subdivision based on the finest mesh and there are 163842 vertices in total. Fig. \ref{fig:Eig_Sph_LM} plots the relative errors in three generalized eigenvalues $\lambda_{400}$,  $\lambda_{500}$ and  $\lambda_{600}$ for a sphere as a function of mesh size $h$. It indicates the convergence rate scales as $h^{2p}$ approximately where $p=3$ when subdivision basis function is employed. As for hand shape object, we start from an initial control mesh with 1442 vertices. Three subsequent subdivisions are processed as well corresponding to 5762, 23042 and 92162 vertices respectively. Again, the reference solution is computed based on one more subdivision with 368642 vertices. The relative error of three generalized eigenvalues $\lambda_{1200}$,  $\lambda_{1300}$ and  $\lambda_{1400}$ is demonstrated in Fig. \ref{fig:Eig_Hand_LM}. From these results, it is apparent that the surface Laplacian equation system can be well discretized by using subdivision basis $\xi({\bf r})$ within IGA framework.

\subsection{EM Scattering from PEC Objects}
Next, several numerical examples are presented to illustrate the efficacy of the proposed GDS-EFIE for analysis of EM scattering problems. In all cases, we present comparison with of radar cross section (RCS) data obtained using the proposed method with that against a well validated Rao-Wilton-Glisson based EFIE code, and against analytical data (when available). 

\begin{figure}[t]
\centering
\includegraphics[width=3.5in]{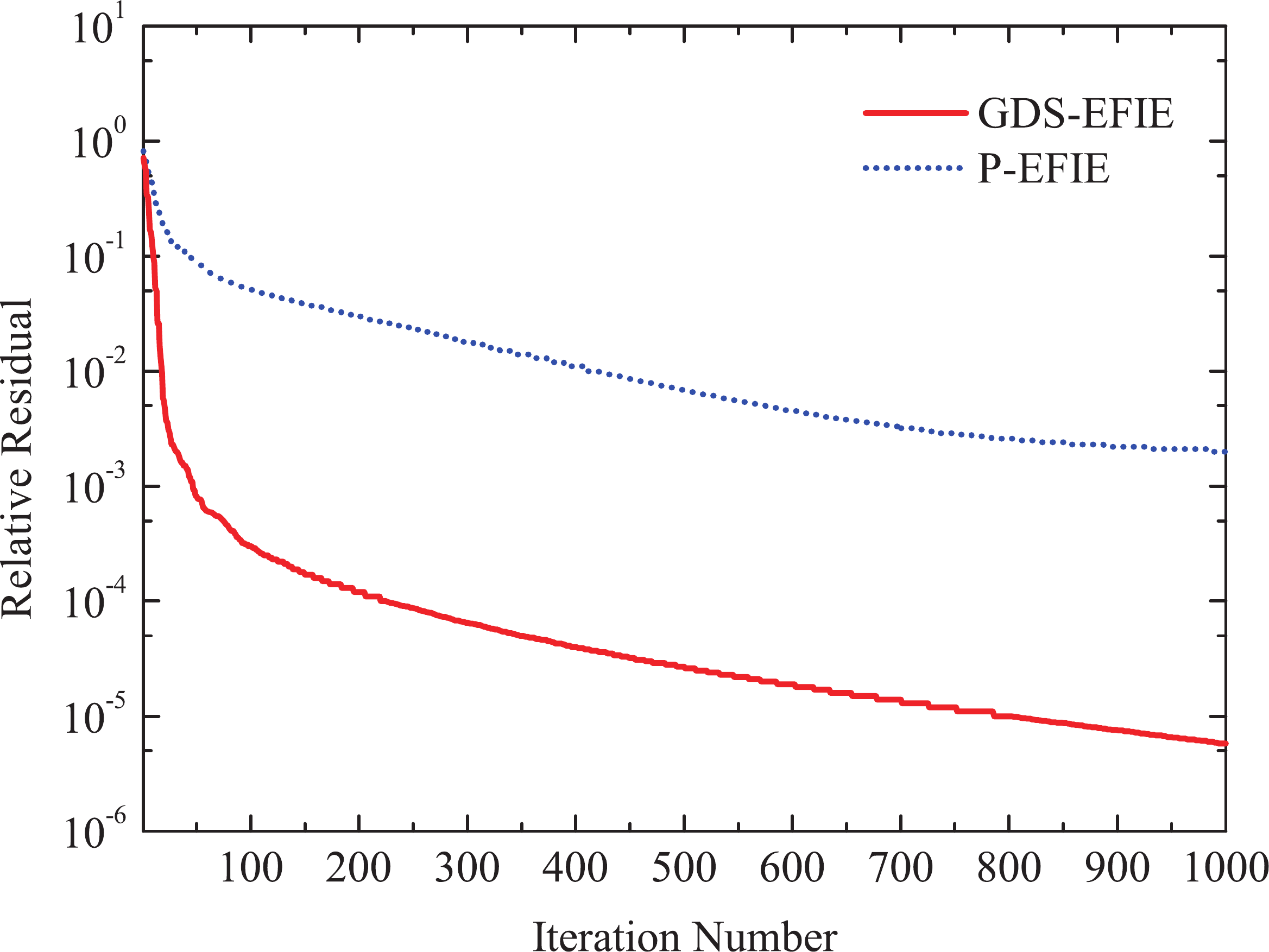}
\caption{Convergence history for a sphere with radius $r=0.67\lambda$ illuminated by a plane wave.}
\label{fig:Sph_Iter}
\end{figure}

First, we consider a PEC sphere with radius $r=0.67\lambda$ illuminated by a plane wave traveling along $-\hat{z}$ with electric field $-\hat{x}$ polarized is considered first. Here $\lambda$ is the working wavelength. There are 642 vertices for discretization. Bistatic RCS solutions from GDS- and P-EFIE \cite{Li2016} are plotted in Fig. \ref{fig:Sph_RCS}.  Both EFIE formulations have an excellent agreement with the reference solution obtained by Mie series method. Fig. \ref{fig:Sph_Iter} depicts the relative residuals when the iterative solver is applied to both GDS- and P-EFIE. With generalized Debye sources as unknowns, GDS-EFIE converges faster compared with P-EFIE.

\begin{figure}[t]
\centering
\includegraphics[width=3.5in]{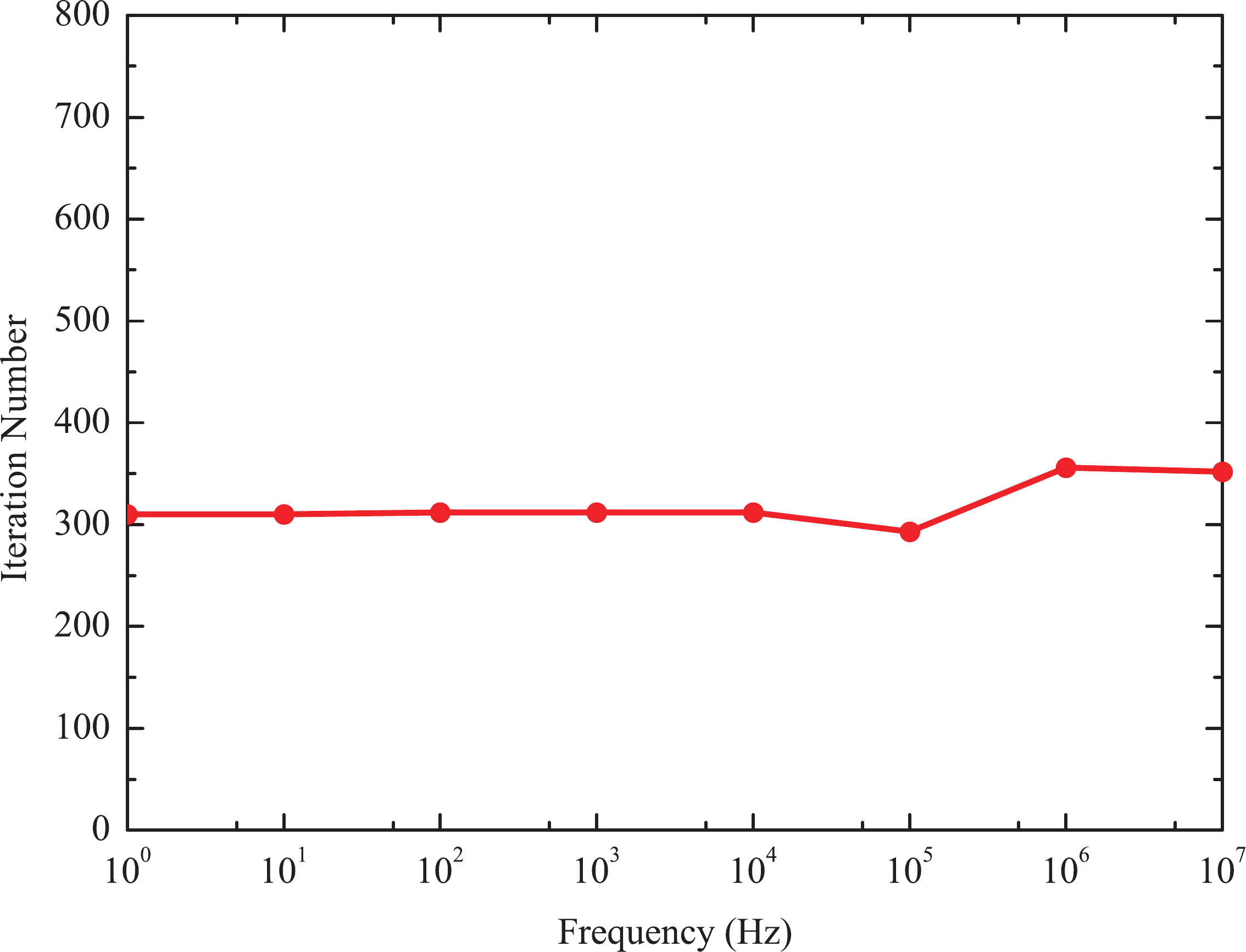}
\caption{Iteration number for GDS-EFIE over a wide band frequency regime.}
\label{fig:Sph_LF_Iter}
\end{figure}

Next, to demonstrate the low-frequency stability of GDS-EFIE, the mesh size is fixed and the working frequency decreases gradually. Fig. \ref{fig:Sph_LF_Iter} plots the iteration numbers to achieve the prescribing tolerance $10^{-5}$ for GDS-EFIE for frequencies ranging from 1Hz to 10MHz. In the figure, the frequency is sampled at every order. We can see that the iteration number is  stable in the low-frequency regime. It is noted that P-EFIE will not converge without any preconditioning techniques. 

\begin{figure}[t]
\centering
\subfigure{}
\includegraphics[width=2.5in]{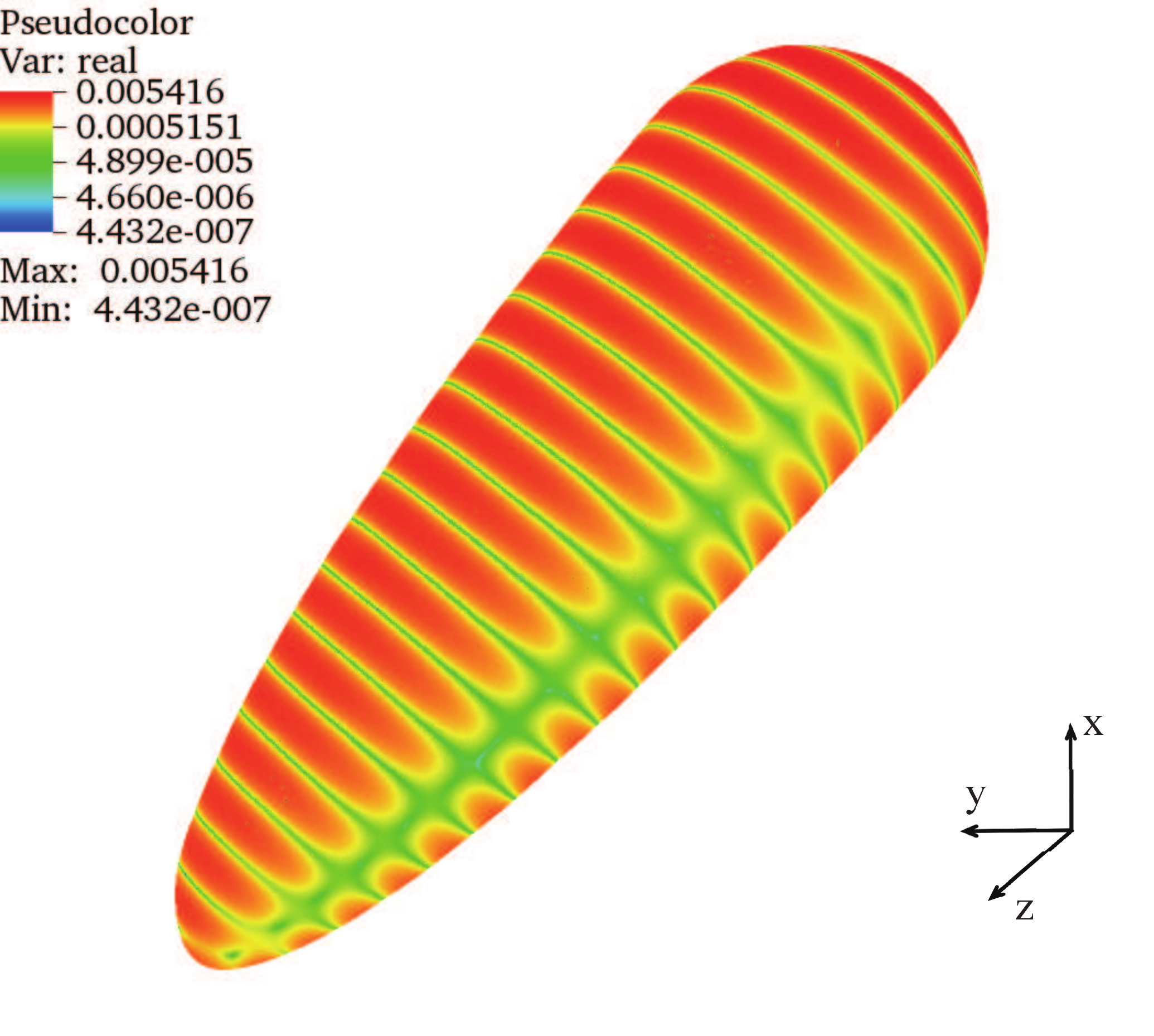}
\subfigure{}
\includegraphics[width=2.5in]{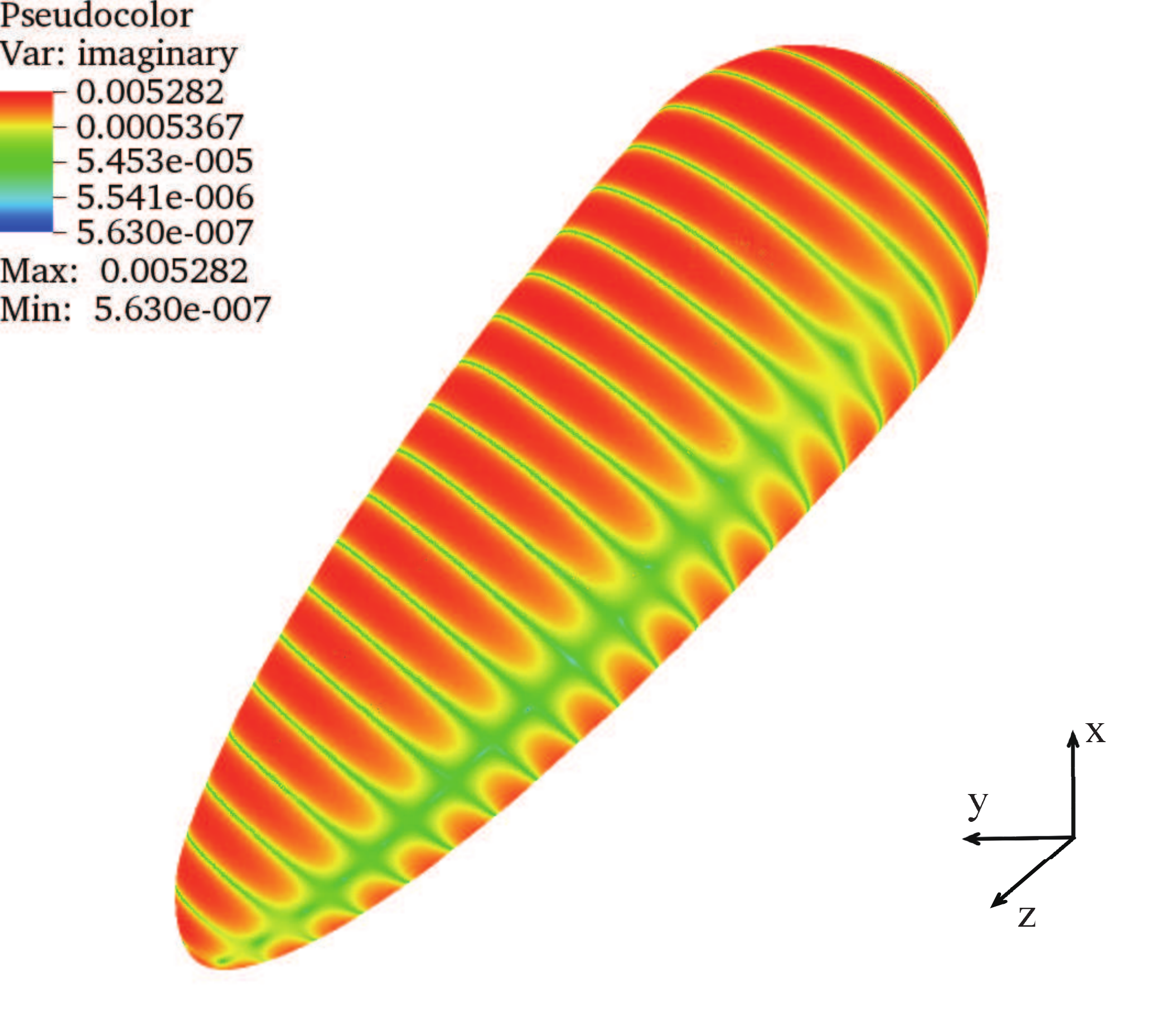}
\caption{Surface current density distribution on a warhead: (a) real part (b) imaginary part.}
\label{fig:Warhead_Cur}
\end{figure}

\begin{figure}[t]
\centering
\includegraphics[width=3.5in]{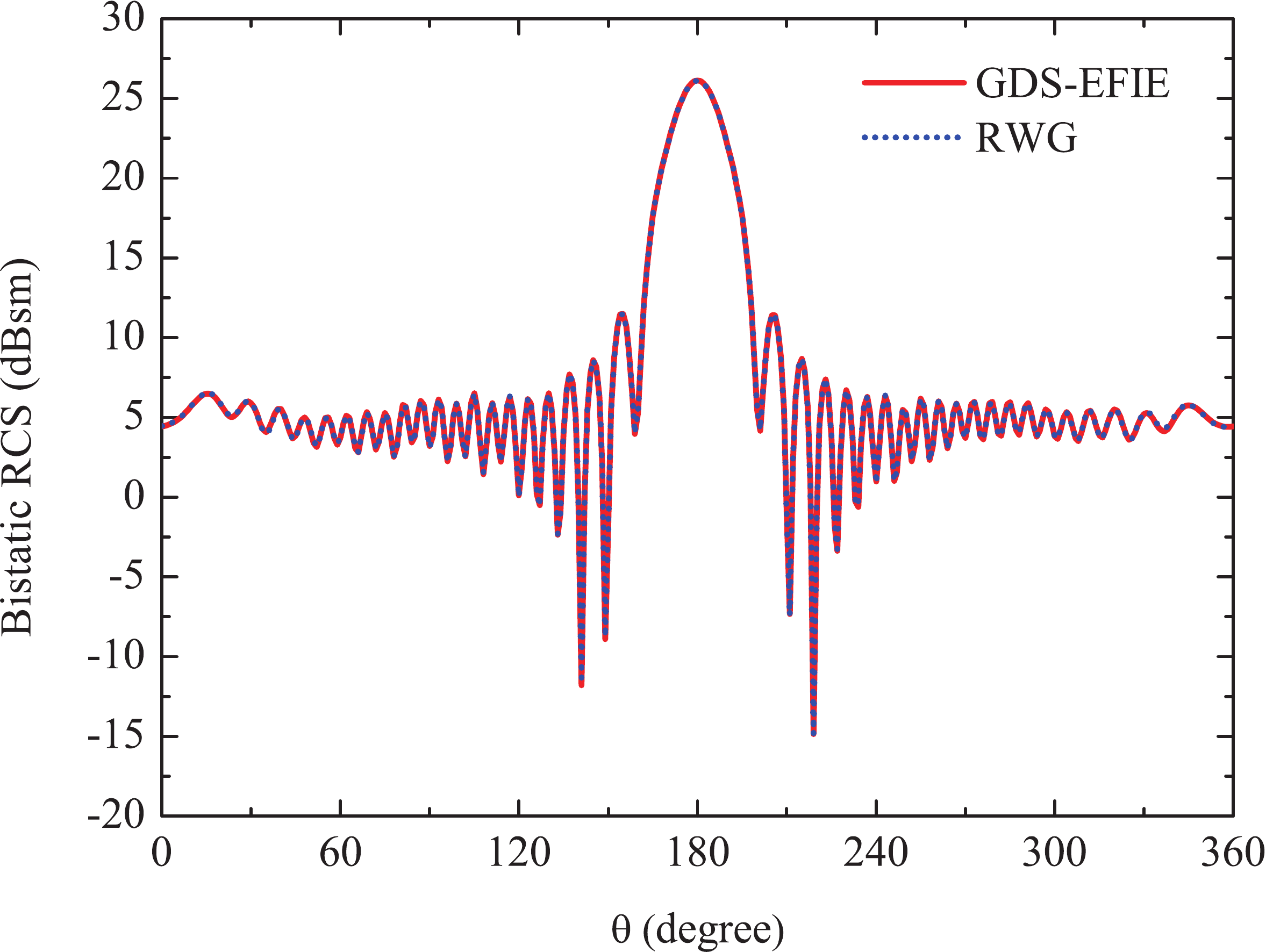}
\caption{Bistatic RCS solutions at $\phi=0$ cut for a warhead.}
\label{fig:Warhead_RCS}
\end{figure}

The second example involves scattering from a warhead which fits into a box $10.4\lambda\times3.5\lambda\times3.5\lambda$. The incident plane wave propagates along $-\hat{z}$ with electric field $-\hat{x}$ polarized. The warhead is discretized by using 21376 triangle elements and then there are 10690 vertices. Fig. \ref{fig:Warhead_Cur} depicts the surface current density of the warhead obtained by GDS-EFIE and it is evident that there are no artificial defects since smooth subdivision basis function is applied. Bistatic solutions from subdivision basis for GDS-EFIE and RWG based EFIE code are plotted in Fig. \ref{fig:Warhead_RCS}. It is evident that both solutions agree with each other very well.


\begin{figure}[t]
\centering
\subfigure{}
\includegraphics[width=2.5in]{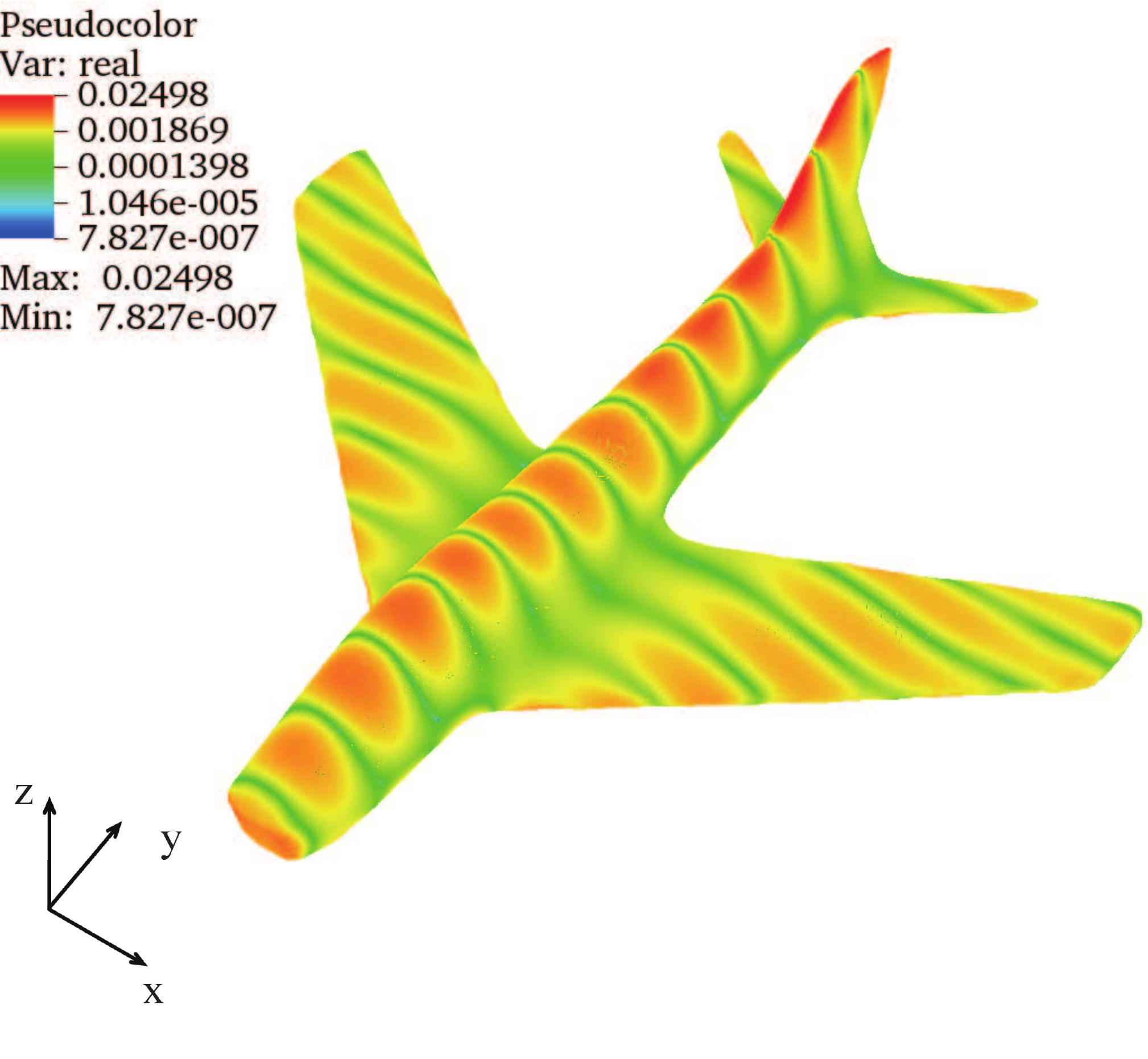}
\subfigure{}
\includegraphics[width=2.5in]{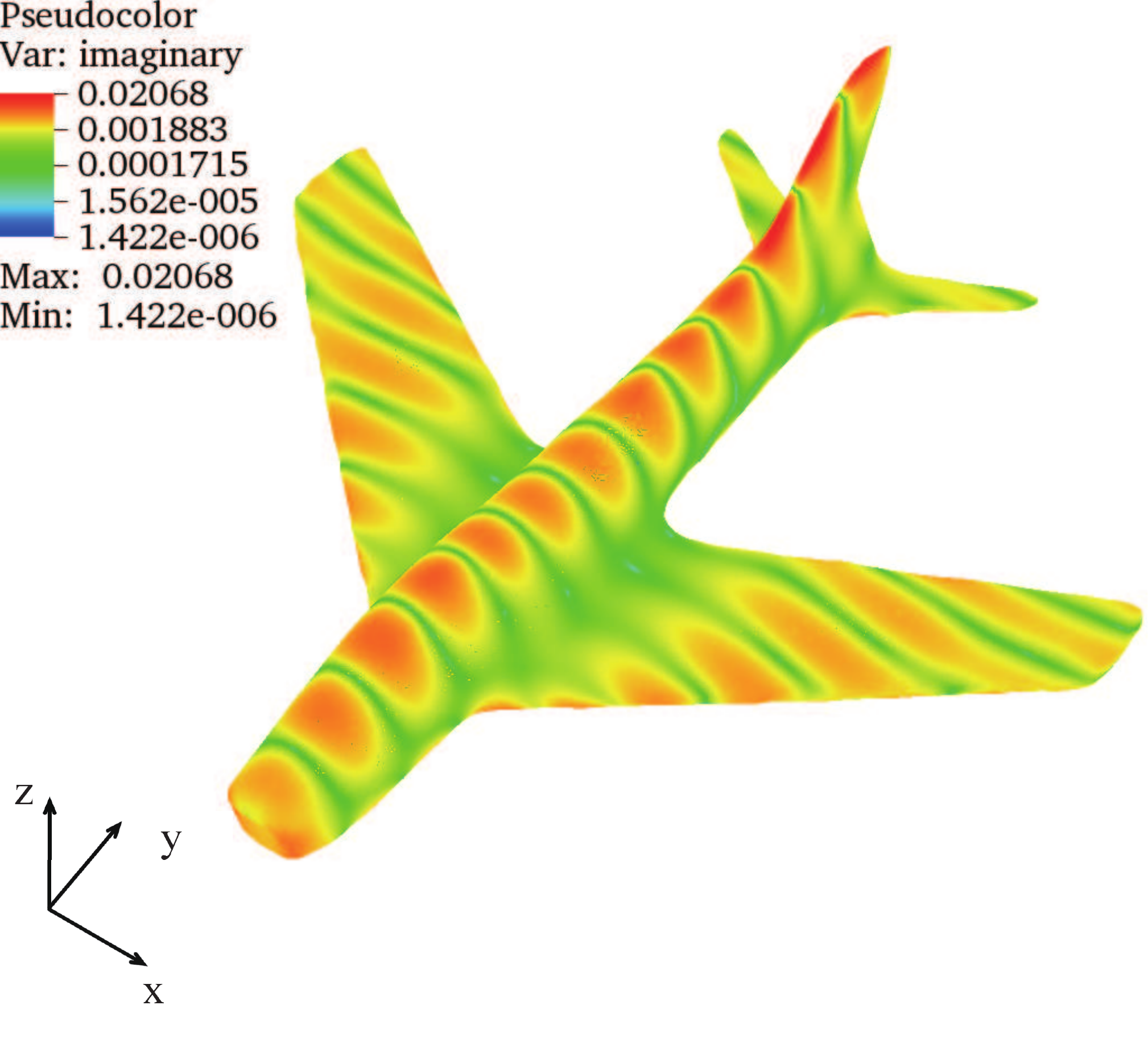}
\caption{Surface current density distribution on a plane model: (a) real part (b) imaginary part.}
\label{fig:Plane_Cur}
\end{figure}

\begin{figure}[t]
\centering
\includegraphics[width=3.5in]{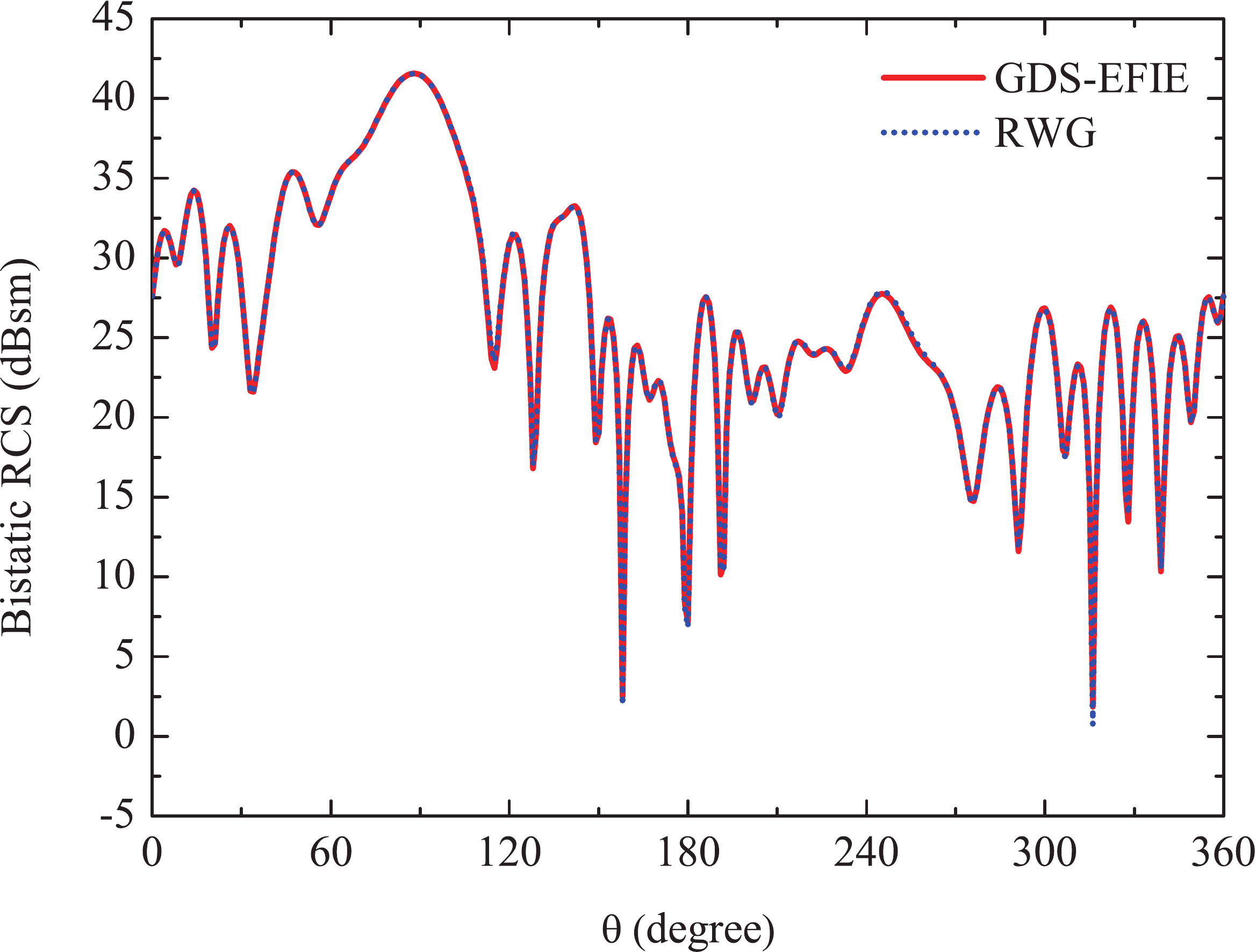}
\caption{Bistatic RCS solutions at $\phi=0$ cut for a plane model.}
\label{fig:Plane_RCS}
\end{figure}

As the last example, scattering from a plane model with electrical size of $6.6\lambda\times6.6\lambda\times1.8\lambda$ is analyzed. The object is illuminated by a plane wave traveling along $\hat{y}$ with electric field $\hat{z}$ polarized. The plane is meshed by 39984 triangle elements associated with 19994 vertices. Fig. \ref{fig:Plane_Cur} shows the surface current density and again, there does not exist any artifacts. Bistatic solutions are plotted in Fig. \ref{fig:Plane_RCS} and solution of GDS-EFIE agrees well with the reference solution from RWG based EFIE. 

\section{Conclusion}
In this work, we have developed an integral formulation called GDS-EFIE based on scalar unknowns by extending the idea of generalized Debye sources to traditional EFIE. The proposed formulation inherits the salient property of the work in \cite{Epstein2010,Chernokozhin2013}, i.e., well-conditioned integral equation at low frequency regime. As is evident in the above description, GDS-EFIE is straightforward to implement numerically. The challenge lies on solving the surface Laplacian equation both effectively and efficiently. Thanks to the smooth subdivision basis, we can find the inverse of Laplace-Beltrami operator within IGA framework with high order accuracy and convergence. Furthermore, no additional continuity constraint is required and our framework can be applied to arbitrarily shaped simply-connected triangular control mesh, without being limited to flat structured mesh. Several numerical examples have been presented to show the well conditioning and flexibility of our proposed GDS-EFIE at both regular and low frequency regimes. Our next steps are to develop a solver for the original Debye source method proposed in \cite{Epstein2010} by using subdivision basis sets within IGA framework, and then extend these to composite/multiply connected objects. 

\section*{Acknowledgment}
This work was supported in part by Hong Kong GRF 712612E, NSFC 61271158, US AOARD 124082 and 134140 contracted through UTAR, and Hong Kong UGC AoE/P04/08. The authors thank the high performance computing center at Michigan State University for computational support.

\bibliographystyle{IEEEtran}
\bibliography{TAP}

\begin{thebibliography}{10}
\providecommand{\url}[1]{#1}
\csname url@samestyle\endcsname
\providecommand{\newblock}{\relax}
\providecommand{\bibinfo}[2]{#2}
\providecommand{\BIBentrySTDinterwordspacing}{\spaceskip=0pt\relax}
\providecommand{\BIBentryALTinterwordstretchfactor}{4}
\providecommand{\BIBentryALTinterwordspacing}{\spaceskip=\fontdimen2\font plus
\BIBentryALTinterwordstretchfactor\fontdimen3\font minus
  \fontdimen4\font\relax}
\providecommand{\BIBforeignlanguage}[2]{{%
\expandafter\ifx\csname l@#1\endcsname\relax
\typeout{** WARNING: IEEEtran.bst: No hyphenation pattern has been}%
\typeout{** loaded for the language `#1'. Using the pattern for}%
\typeout{** the default language instead.}%
\else
\language=\csname l@#1\endcsname
\fi
#2}}
\providecommand{\BIBdecl}{\relax}
\BIBdecl

\bibitem{poggio1970}
A.~J. Poggio and E.~K. Miller, \emph{Integral Equation Solutions of
  Three-Dimensional Scattering Problems, Computer Techniques in
  Electromagnetics, edited by R. Mittra}.\hskip 1em plus 0.5em minus
  0.4em\relax Oxford, United Kingdom: Pergamon Press, 1973, pp. 159-261.

\bibitem{peterson1998}
A.~Peterson, S.~Ray, and R.~Mittra, \emph{Computational Methods for
  Electromagnetics}.\hskip 1em plus 0.5em minus 0.4em\relax New York, NY, USA:
  IEEE Press, 1998.

\bibitem{jin2011}
J.-M. Jin, \emph{Theory and Computation of Electromagnetic Fields, 2nd
  ed.}\hskip 1em plus 0.5em minus 0.4em\relax Hoboken, NJ, USA: Wiley, 2015.

\bibitem{volakis}
J.~L. Volakis and K.~Sertel, \emph{Integral Equation Methods for
  Electromagnetics}.\hskip 1em plus 0.5em minus 0.4em\relax West Perth, WA,
  Australia: SciTech, 2012.

\bibitem{kolundvzija2002}
B.~M. Kolund{\v{z}}ija and A.~R. Djordjevi{\'c}, \emph{Electromagnetic Modeling
  of Composite Metallic and Dielectric Structures}.\hskip 1em plus 0.5em minus
  0.4em\relax Norwood, MA, USA: Artech House, 2002.

\bibitem{Wilton1981}
D.~R. Wilton and A.~W. Glisson, ``On improving the electric field integral
  equation at low frequencies,'' in \emph{Proceedings of URSI Radio Science
  Meeting Digest}, June 1981, p.~24.

\bibitem{Wu1995}
W.~Wu, A.~W. Glisson, and D.~Kajfez, ``A study of two numerical solution
  procedures for electric field integral equation at low frequency,''
  \emph{Applied Computational Electromagnetics Society Journal}, vol.~10, pp.
  69--80, Mar 1995.

\bibitem{Burton1995}
M.~Burton and S.~Kashyap, ``A study of recent moment-method algorithm that is
  accurate to very low frequencies,'' \emph{Applied Computational
  Electromagnetics Society Journal}, vol.~10, pp. 58--60, Mar 1995.

\bibitem{Taskinen2006}
M.~Taskinen and P.~Yl{\"{a}}-Oijala, ``{Current and charge integral equation
  formulation},'' \emph{IEEE Trans. Antennas Propag.}, vol.~54, no.~1, pp.
  58--67, 2006.

\bibitem{graglia2015}
R.~Graglia and A.~F. Peterson, \emph{Higher-Order Techniques in Computational
  Electromagnetics}.\hskip 1em plus 0.5em minus 0.4em\relax Edison, NJ, USA:
  SciTech, 2015.

\bibitem{warnick2008}
K.~Warnick, \emph{Numerical Analysis for Electromagnetic Integral
  Equations}.\hskip 1em plus 0.5em minus 0.4em\relax Norwood, MA, USA: Artech
  House, 2008.

\bibitem{peterson2006}
A.~Peterson, \emph{Mapped Vector Basis Functions for Electromagnetic Integral
  Equations}.\hskip 1em plus 0.5em minus 0.4em\relax San Rafael, CA, USA:
  Morgan \& Claypool, 2006.

\bibitem{Qian2010}
Z.~G. Qian and W.~C. Chew, ``{Enhanced A-EFIE with perturbation method},''
  \emph{IEEE Trans. Antennas Propag.}, vol.~58, no.~10, pp. 3256--3264, 2010.

\bibitem{Andriulli2008}
F.~P. Andriulli, K.~Cools, H.~Baǧci, F.~Olyslager, A.~Buffa, S.~Christiansen,
  and E.~Michielssen, ``{A multiplicative Calder\'{o}n preconditioner for the
  electric field integral equation},'' \emph{IEEE Trans. Antennas Propag.},
  vol.~56, no.~8, pp. 2398--2412, 2008.

\bibitem{Vipiana2005}
F.~Vipiana, P.~Pirinoli, and G.~Vecchi, ``{A multiresolution method of moments
  for triangular meshes},'' \emph{IEEE Trans. Antennas Propag.}, vol.~53,
  no.~7, pp. 2247--2258, 2005.

\bibitem{Vecchi1999}
G.~Vecchi, ``Loop-star decomposition of basis functions in the discretization
  of the {EFIE},'' \emph{IEEE Transactions on Antennas and Propagation},
  vol.~47, no.~2, pp. 339--346, Feb 1999.

\bibitem{Zhao2000}
J.~S. Zhao and W.~C. Chew, ``{Integral equation solution of Maxwell's equations
  from zero frequency to microwave frequencies},'' \emph{IEEE Trans. Antennas
  Propag.}, vol.~48, no.~10, pp. 1635--1645, 2000.

\bibitem{Xiong2013}
X.~Y.~Z. Xiong, L.~J. Jiang, W.~E.~I. Sha, and Y.-H. Lo, ``A new {EFIE} method
  based on coulomb gauge for the low-frequency electromagnetic analysis,''
  \emph{Prog. Electromagn. Res.}, vol. 140, no. April, pp. 613--631, 2013.

\bibitem{Epstein2010}
C.~Epstein and L.~Greengard, ``{Debye sources and the numerical solution of the
  time harmonic {Maxwell} equations},'' \emph{Commun. Pure Appl. Math.},
  vol.~63, no.~4, pp. 413--463, 2010.

\bibitem{Epstein2013a}
C.~L. Epstein, L.~Greengard, and M.~O'Neil, ``Debye sources and the numerical
  solution of the time harmonic {Maxwell} equations ii,'' \emph{Commun. Pure
  Appl. Math.}, vol.~66, no.~5, pp. 753--789, 2013.

\bibitem{Epstein2015}
------, ``Debye sources, {Beltrami} fields, and a complex structure on
  {Maxwell} fields,'' \emph{Commun. Pure Appl. Math.}, vol.~68, no.~12, pp.
  2237--2280, 2015.

\bibitem{Vico2014}
F.~Vico, M.~Ferrando, L.~Greengard, and Z.~Gimbutas, ``The decoupled potential
  integral equation for time-harmonic electromagnetic scattering,'' \emph{arXiv
  preprint, arXiv:1404.0749}, 2014.

\bibitem{Vico2015}
F.~Vico, M.~Ferrando-Bataller, T.~Jim{\'{e}}nez, and A.~Berenguer, ``A high
  order locally corrected nystr{\"{o}}m implementation of the decoupled
  potential integral equation,'' in \emph{2015 9th European Conference on
  Antennas and Propagation (EuCAP)}, May 2015, pp. 1--4.

\bibitem{ChoChew2014}
W.~{Cho Chew}, ``Vector potential electromagnetics with generalized gauge for
  inhomogeneous media: Formulation,'' \emph{Prog. Electromagn. Res.}, vol. 149,
  no. August, pp. 69--84, 2014.

\bibitem{Liu2015}
Q.~S. Liu, S.~Sun, and W.~C. Chew, ``An integral equation method based on
  vector and scalar potential formulations,'' in \emph{2015 IEEE International
  Symposium on Antennas and Propagation USNC/URSI National Radio Science
  Meeting}, July 2015, pp. 744--745.

\bibitem{Chernokozhin2013}
E.~V. Chernokozhin and A.~Boag, ``{Method of generalized {Debye} sources for
  the analysis of electromagnetic scattering by perfectly conducting bodies
  with piecewise smooth boundaries},'' \emph{IEEE Trans. Antennas Propag.},
  vol.~61, no.~4, pp. 2108--2115, 2013.

\bibitem{Li2016}
J.~Li, D.~Dault, B.~Liu, Y.~Tong, and B.~Shanker, ``{Subdivision based
  isogeometric analysis technique for electric field integral equations for
  simply connected structures},'' \emph{J. Comput. Phys.}, vol. 319, pp.
  145--162, 2016.

\bibitem{Hughes2005}
T.~J.~R. Hughes, J.~A. Cottrell, and Y.~Bazilevs, ``{Isogeometric analysis:
  CAD, finite elements, NURBS, exact geometry and mesh refinement},''
  \emph{Comput. Methods Appl. Mech. Eng.}, vol. 194, no. 39-41, pp. 4135--4195,
  2005.

\bibitem{Loop1987}
C.~Loop, ``{Smooth subdivision surfaces based on triangles},'' Master's thesis,
  University of Utah, 1987.

\bibitem{Rao1982}
S.~Rao, D.~Wilton, and A.~Glisson, ``{Electromagnetic scattering by surfaces of
  arbitrary shape},'' \emph{IEEE Trans. Antennas Propag.}, vol.~30, no.~3, pp.
  409--418, 1982.

\bibitem{Raviart1977}
P.~A. Raviart and J.~M. Thomas, \emph{A Mixed Finite Element Method for 2-nd
  Order Elliptic Problems, Mathematical Aspects of the Finite Element Method,
  Lecture notes in Mathematics}.\hskip 1em plus 0.5em minus 0.4em\relax New
  York, NY, USA: Springer-Verlag, 1977, vol. 606, pp. 292--315.

\bibitem{Nair2011}
N.~V. Nair and B.~Shanker, ``{Generalized method of moments: A novel
  discretization technique for integral equations},'' \emph{IEEE Trans.
  Antennas Propag.}, vol.~59, no.~6, pp. 2280--2293, 2011.

\bibitem{Dault2014}
D.~L. Dault, N.~V. Nair, J.~Li, and B.~Shanker, ``The generalized method of
  moments for electromagnetic boundary integral equations,'' \emph{Antennas
  Propagation, IEEE Trans.}, vol.~62, no.~6, pp. 3174--3188, 2014.

\bibitem{Peng2013}
Z.~Peng, K.~H. Lim, and J.~F. Lee, ``A discontinuous galerkin surface integral
  equation method for electromagnetic wave scattering from nonpenetrable
  targets,'' \emph{IEEE Transactions on Antennas and Propagation}, vol.~61,
  no.~7, pp. 3617--3628, July 2013.

\bibitem{Bazilevs2010}
Y.~Bazilevs, V.~M. Calo, J.~A. Cottrell, J.~A. Evans, T.~J.~R. Hughes,
  S.~Lipton, M.~A. Scott, and T.~W. Sederberg, ``{Isogeometric analysis using
  T-splines},'' \emph{Comput. Methods Appl. Mech. Eng.}, vol. 199, no. 5-8, pp.
  229--263, 2010.

\bibitem{Stam1998}
J.~Stam, ``{Evaluation of Loop subdivision surface},'' in \emph{Computer
  Graphics Proceedings ACM SIGGRAPH}, 1998.

\bibitem{Juttler2016}
B.~J{\"{u}}ttler, A.~Mantzaflaris, R.~Perl, and M.~Rumpf, ``{On numerical
  integration in isogeometric subdivision methods for PDEs on surfaces},''
  \emph{Comput. Methods Appl. Mech. Eng.}, vol. 302, pp. 131--146, 2016.

\end{thebibliography}






\end{document}